\documentclass[10pt,reqno]{amsart}
\usepackage{amsmath}
\usepackage{amsfonts,amssymb}
\def\EQ{\begin{equation}}
\def\EQs{\begin{eqnarray}}
\def\endEQ{\end{equation}}
\def\endEQs{\end{eqnarray}}

\def\eqref#1{(\ref{#1})}
\def\eqrefs#1#2{(\ref{#1}) and~(\ref{#2})}
\def\eqsref#1#2{(\ref{#1})--(\ref{#2})}

\def\mixedindices#1#2{{\mathstrut}^{#1}_{#2}}
\def\downindex#1{{\mathstrut}_{#1}}
\def\upindex#1{{\mathstrut}^{#1}}

\def\der#1{\partial\downindex{#1}}
\def\Dx{D\downindex{x}}
\def\Dt{D\downindex{t}}
\def\Dinvx{D\mixedindices{-1}{x}}

\def\covder#1{\nabla\downindex{#1}}

\def\parder#1{\partial/\partial{#1}}

\def\map{\gamma}
\def\mapder#1{\map_{#1}}
\def\cmapder#1{\bar\map_{#1}}
\def\mapcurv{\kappa}

\def\hperp{h_\perp}
\def\hpar{h_\parallel}
\def\w{\varpi}

\def\scalcurv{\chi}
\def\riem{R}
\def\tors{T}

\def\gconx{{}^\g\omega}

\def\e{e}
\def\duale{e^*}
\def\conx{\omega}

\def\frder{{\mathcal D}}
\def\frtors{{\mathfrak T}}
\def\frcurv{{\mathfrak R}}

\def\ehook#1{\e\downindex{#1}}
\def\conxhook#1{\conx\downindex{#1}}
\def\frderhook#1{{\mathcal D}\downindex{#1}}
\def\frcurvhook#1{\frcurv\downindex{#1}}
\def\frtorshook#1{\frtors\downindex{#1}}

\def\mvec#1{{\bf m}^{#1}}
\def\hvec#1{{\bf h}^{#1}}

\def\c#1#2{c^{#1}{}_{#2}}

\def\Rop{{\mathcal R}}
\def\Jop{{\mathcal J}}
\def\Hop{{\mathcal H}}

\def\Eop{{\mathcal E}}

\def\Kop{{\mathcal K}}

\def\ham#1{{\mathfrak #1}}
\def\Ham{H}

\def\adRop{{\Rop^*}}

\def\sympform{{\boldsymbol \omega}}

\def\u{u}
\def\uni{\theta}
\def\aduni{\tilde\theta}
\def\adsquni{\underset{\textstyle{\tilde\ }}{\theta}}
\def\adu{\tilde\u}
\def\v{v}
\def\q{q}

\def\hook{\rfloor}

\def\op{{\mathcal D}}
\def\univar{\delta_{\uni}}
\def\uniopvar#1{\delta_{#1(\uni)}}

\def\alg#1{{\mathfrak {#1}}}
\def\g{\alg{g}}
\def\h{\alg{h}}
\def\m{\alg{m}}
\def\a{\alg{a}}

\def\cen{\alg{c}}

\def\hcen{\cen(\ehook{X})_\h}
\def\mcen{\cen(\ehook{X})_\m}
\def\hcenperp{\hcen^\perp}
\def\mcenperp{\mcen^\perp}

\def\gcen{\cen(\ehook{X})_\g}
\def\gcenperp{\gcen^\perp}

\def\isoH{H{}^*_{X}}
\def\equivH{H^*_\parallel}

\def\ad{{\rm ad}}
\def\Ad{{\rm Ad}}
\def\tr{{\rm tr}}
\def\adsq{{\mathcal X}}

\def\grpid{{\rm e}}

\def\J{J}

\def\id{{\rm id}}
\def\o{o}

\def\ie/{i.e.}
\def\eg/{e.g.}
\def\etc/{etc.}
\def\const{{\rm const}}

\def\Rnum{{\mathbb R}}
\def\Cnum{{\mathbb C}}

\def\i{{\rm i}}
\def\supth#1{\upindex{(#1)}}
\def\nth#1{{(#1)}}
\def\d{{\rm d}}
\def\pr{{\rm pr}}

\def\inv{{}^{-1}}

\def\th{{}^{\rm th}}

\begin{document}

\title[Soliton equations and bi-Hamiltonian curve flows]{
Group-invariant soliton equations and\\
bi-Hamiltonian geometric curve flows in\\ 
Riemannian symmetric spaces } 

\author{Stephen C. Anco}
\address{Department of Mathematics, Brock University, Canada}
\email{sanco@brocku.ca}

\begin{abstract}
Universal bi-Hamiltonian hierarchies of group-invariant 
(multicomponent) soliton equations 
are derived from non-stretching geometric curve flows $\map(t,x)$
in Riemannian symmetric spaces $M=G/H$,
including compact semisimple Lie groups $M=K$
for $G=K\times K$, $H={\rm diag}\ G$. 
The derivation of these soliton hierarchies 
utilizes a moving parallel frame and connection 1-form 
along the curve flows, related to the Klein geometry 
of the Lie group $G\supset H$
where $H$ is the local frame structure group. 
The soliton equations arise in explicit form 
from the induced flow on the frame components of
the principal normal vector $N=\covder{x}\mapder{x}$ along each curve,
and display invariance under the equivalence subgroup
in $H$ that preserves the unit tangent vector $T=\mapder{x}$ in the framing 
at any point $x$ on a curve. 
Their bi-Hamiltonian integrability structure is shown to be geometrically
encoded in the Cartan structure equations for torsion and curvature 
of the parallel frame and its connection 1-form in the tangent space 
$T_\map M$ of the curve flow. 
The hierarchies include group-invariant versions of 
sine-Gordon (SG) and modified Korteweg-de Vries (mKdV) soliton equations 
that are found to be universally given by curve flows describing
non-stretching wave maps and mKdV analogs of non-stretching Schrodinger maps 
on $G/H$.
These results provide a geometric interpretation 
and explicit bi-Hamiltonian formulation 
for many known multicomponent soliton equations. 
Moreover, all examples of group-invariant (multicomponent) soliton equations
given by the present geometric framework 
can be constructed in an explicit fashion based on 
Cartan's classification of symmetric spaces. 
\end{abstract}

\maketitle

\section{Introduction and Overview}

The theory of integrable soliton equations displays many deep links
to differential geometry, particularly as found in the study of
geometric curve flows. 
In this paper, 
group-invariant soliton equations 
and their bi-Hamiltonian integrability structure
are derived from studying non-stretching flows of curves
in symmetric spaces $G/H$.
Such spaces describe curved $G$-invariant Riemannian manifolds,
generalizing the classical two-dimensional Riemannian geometries \ie/,
the sphere $S^2 \simeq SO(3)/SO(2)$, 
the hyperbolic plane $H^2 \simeq SL(2,\Rnum)/SO(2)$,
and the Euclidean plane $\Rnum^2 \simeq Euc(2)/SO(2)$,
which are characterized by constant (positive, negative, zero) curvature.

Geometric curve flows in constant curvature spaces $M=S^2,H^2,\Rnum^2$
are well known 
to yield 
\cite{DoliwaSantini,GoldsteinPetrich,LangerPerline,Gurses}
the mKdV hierarchy of scalar soliton equations
and its hereditary recursion operator. 
The starting point is to formulate the flow of 
a non-stretching (inextensible) curve
$\map(t,x)$ 
as geometrically given by 
$\mapder{t} = \hperp N + \hpar T$
in an adapted moving frame $(T,N)=(\mapder{x},*\mapder{x})$
(\ie/ a Frenet frame) 
along the curve,
where $x$ is arclength
and $*$ is the Hodge-star operator 
on vectors in the tangent plane $\Rnum^2\simeq T_x M$.
Preservation of the non-stretching condition 
$\mapder{x}\cdot\mapder{x}=1$
under the curve flow yields the relation 
$\der{x}\hpar=\mapcurv \hperp$,
allowing the tangential component $\hpar$ of the flow
to be expressed in terms of the normal component $\hperp$
and the curvature invariant 
$\mapcurv=N\cdot\covder{x}T$ 
of the curve. 
The flow equation combined with the Serret-Frenet structure equations
of the moving frame then determine an evolution of this curvature invariant
$\mapcurv_{t} = \Rop(\hperp) + \scalcurv\hperp$
where $\Rop = \Dx^2 +\mapcurv^2 +\mapcurv_{x}\Dinvx\mapcurv$ 
is seen to be 
the hereditary recursion operator of 
the mKdV hierarchy of soliton equations \cite{Olver},
and $\scalcurv$ is the Gaussian curvature $(+1,-1,0)$ of $M$.
In particular, 
the flow given by 
$\hperp=\mapcurv_{x}$, $\hpar=\frac{1}{2}\mapcurv^2$
yields the mKdV evolution equation on $\mapcurv$
to within a convective term,
$\mapcurv_{t} -\scalcurv\mapcurv_{x}=
\mapcurv_{3x} +\frac{3}{2}\mapcurv^2 \mapcurv_{x}$
(called the $+1$ flow in the hierarchy).
Higher order mKdV evolution equations arise from
linear combinations of flows 
$\hperp=\Rop^{n}(\mapcurv_{x})$,
$\hpar=\Dinvx(\mapcurv\Rop^{n}(\mapcurv_{x}))$,
$n=1,2,\ldots$.
The entire mKdV hierarchy 
$\mapcurv_{t} -\scalcurv\mapcurv_{x}=\Rop^n(\mapcurv_{x})$
thus sits in the class of geometric flows
in which the evolution of the curve in the normal direction 
is a function $\hperp=\hperp(\mapcurv,\mapcurv_{x},\ldots)$
of the differential invariants of the curve
\cite{Maribeffa1,Maribeffa2}.
Such flows correspond to geometric equations 
$\mapder{t} =f(\mapder{x},\covder{x}\mapder{x},\ldots)$
satisfied by the curve $\map(t,x)$.

It is less widely known that the geometric curve equations
produced in this manner from the mKdV hierarchy of evolutions on $\mapcurv$
are close analogs of Schrodinger maps 
$\mapder{t} =J(\mapder{x})$, 
$J=\i\covder{x}$, on $S^2 \simeq \Cnum P^1$
(identifying $\i$ with $*$
after complexification of the tangent plane).
For instance the mKdV evolution itself yields
$\mapder{t} =\mapcurv_{x} N +\frac{1}{2}\mapcurv^2 T
= \covder{x}(\mapcurv\; {*\mapder{x}}) +\frac{3}{2}\mapcurv^2 \mapder{x}$
describing \cite{Anco1} a geometric curve equation
$\mapder{t} =K_\map(\mapder{x})$, 
$K_\map=\covder{x}^2+\frac{3}{2}|\covder{x}\mapder{x}|^2$
as obtained through the Serret-Frenet equations
$\covder{x}T =\mapcurv N$, $\covder{x}N =-\mapcurv T$. 
This flow operator $K_\map$ is the mKdV analog of 
a modified Schrodinger operator 
$J_\map = \i\covder{x} + (\arg \mapder{x})_x$
that preserves $|\mapder{x}|=1$. 
One geometrical difference between these operators is that 
$\cmapder{x}\cdot J_\map(\mapder{x})=0$
whereas $\cmapder{x}\cdot K_\map(\mapder{x})=\frac{1}{2} \mapcurv^2 >0$,
implying (by dimensional considerations) that 
the non-stretching Schrodinger flow of the curve $\map(t,x)$ in $S^2$
is actually stationary
$\mapder{t} =J_\map(\mapder{x})=0$
(so thus $\mapcurv_{t}=0$),
while in contrast the non-stretching mKdV flow 
$\mapder{t} =K_\map(\mapder{x})$
is dynamical.   

A related geometric evolution \cite{Anco1} 
comes from the kernel of the mKdV recursion operator $\Rop(\hperp)=0$
giving the evolution equation
$\mapcurv_{t}= \scalcurv \hperp$
(called the $-1$ flow in the mKdV hierarchy),
with $\hperp$ satisfying
$\der{x}\hperp = -\mapcurv\hpar$, 
$\der{x}\hpar = \mapcurv\hperp$.
This flow obeys the conservation law
$0=\der{x}( \hpar^2 +\hperp^2 ) = \covder{x}(\mapder{t}\cdot\mapder{t})$
which implies it is conformally equivalent 
(under rescalings of $t$) to a flow with uniform speed
$|\mapder{t}|=\sqrt{\hpar^2 +\hperp^2}=1$.
The evolution of the curve is then given by 
$\hperp= \sin(\theta)$, $\hpar = \cos(\theta)$
in terms of the nonlocal invariant 
$\theta=-\int\mapcurv dx$ of $\map$,
yielding the sine-Gordon equation 
$\theta_{tx}= -\scalcurv \sin(\theta)$
induced from the evolution on $\mapcurv$
\cite{Lamb,NakayamaSegurWaditi}.
Equivalently, the relations
$\theta =\arcsin(\mapcurv_{t}/\scalcurv)$
and $\theta_{x}=-\mapcurv$
yield a hyperbolic equation 
$\mapcurv_{tx}=-\mapcurv\sqrt{\scalcurv^2-\mapcurv_{t}^2}$
on $\mapcurv$.
The corresponding geometric curve equation satisfied by $\map(t,x)$ is
$\covder{x}\mapder{t} = 
\covder{x}(\sin(\theta) N) + \covder{x}(\cos(\theta) T)
=0$
after the Serret-Frenet equations are used.
This curve flow is recognized \cite{Anco1} to be 
a uniform speed, non-stretching wave map 
$\covder{x}\mapder{t} = \covder{t}\mapder{x} = 0$
on $S^2$,
with $t,x$ viewed as light cone coordinates for the wave operator. 

Moreover, 
the geometry of the two-dimensional surfaces swept out by
all these non-stretching curve flows $\map(t,x)$ in $M=S^2,H^2,\Rnum^2$
turns out to encode the bi-Hamiltonian integrability structure 
\cite{SandersWang1}
of the mKdV soliton hierarchy.
The relevant geometry \cite{Anco1}
is given by the Riemannian connections
$\covder{t},\covder{x}$ in the flow direction $\mapder{t}$
and tangent direction $\mapder{x}$,
specifically that they have vanishing torsion
$\covder{t}\mapder{x} - \covder{x}\mapder{t} = [\mapder{t},\mapder{x}]
=0$
and carry constant curvature
$[\covder{t},\covder{x}]=\riem(\mapder{t},\mapder{x})
=\scalcurv (\mapder{t}\cdot*\mapder{x})*$
where $\riem(X,Y)$ is the Riemann curvature tensor of $M$
with $X,Y$ in the tangent plane. 
A projection of these equations into the normal space of $\map$ yields
$\mapcurv_{t}= \Hop(\w)+\scalcurv\hperp$, 
$\w=\Jop(\hperp)$,
expressed in terms of the flow invariant
$\w=N\cdot\covder{t}T$, 
where
$\Hop=\Dx$,
$\Jop=\Dx +\mapcurv\Dinvx\mapcurv$
are found to be respectively 
the mKdV Hamiltonian cosymplectic and symplectic operators
for the hierarchy of mKdV evolutions on $\mapcurv$.
In this setting 
$\hperp \parder{\mapcurv}$ represents a Hamiltonian vector field,
and $\w \d\mapcurv$ represents a variational covector field
with $\w=\delta\Ham/\delta\mapcurv$
holding for some Hamiltonian $\Ham=\Ham(\mapcurv,\mapcurv_{x},\ldots)$.
Compatibility of the operators $\Hop,\Jop$ 
implies that $\Rop=\Hop\circ\Jop$ and its adjoint $\adRop=\Jop\circ\Hop$
generate a hierarchy of 
commuting vector fields
$\hperp\supth{n} = \Rop^n(\mapcurv_{x})$
and corresponding involutive covector fields
$\w\supth{n} = \adRop^n(\mapcurv)$, 
$n=0,1,2,\ldots$, 
related by 
$\hperp\supth{n} =\Hop(\w\supth{n})$,
$\w\supth{n+1}=\Jop(\hperp\supth{n})$.
For $n\ge 1$ this hierarchy possesses 
a bi-Hamiltonian structure
$\hperp\supth{n} 
=\Hop(\delta\Ham\supth{n}/\delta\mapcurv)
=\Jop\inv(\delta\Ham\supth{n+1}/\delta\mapcurv)$
where $\Jop\inv$ is the formal inverse of $\Jop$
defined on the $x$-jet space of $\mapcurv$,
and with the local Hamiltonians 
$\Ham\supth{n}=\Ham\supth{n}(\mapcurv,\mapcurv_{x},\ldots)$
determined by
$\w\supth{n}=\delta\Ham\supth{n}/\delta\mapcurv$.
Thus $\Hop,\Jop\inv$ are formally a bi-Hamiltonian pair of 
cosymplectic operators.
Alternatively, 
$\Eop = \Hop\Jop\Hop
= \Dx^3 +\mapcurv^2\Dx +\mapcurv_{x}\Dinvx(\mapcurv\Dx)$
provides an explicit cosymplectic operator 
compatible with $\Hop$, giving the bi-Hamiltonian structure 
$\hperp\supth{n} 
=\Eop(\delta\Ham\supth{n-1}/\delta\mapcurv)
=\Hop(\delta\Ham\supth{n}/\delta\mapcurv)$.

The bottom of the mKdV hierarchy 
$\hperp\supth{0}=\mapcurv_{x}$, 
$\w\supth{0}=\mapcurv$,
$\Ham\supth{0}=\frac{1}{2}\mapcurv^2$
originates geometrically from the $x$-translation vector field
$\mapcurv_{x}\parder{\mapcurv}$
which is a symmetry of the operators $\Hop,\Jop$.
In terms of the jet space variables $(x,\mapcurv,\mapcurv_{x},\ldots)$,
the entire hierarchy has the form of 
homogeneous polynomials with non-zero scaling weight
under the mKdV scaling symmetry 
$x\rightarrow \lambda x$, $\mapcurv\rightarrow \lambda^{-1} \mapcurv$.
As a consequence, through a scaling formula
(cf. \cite{Anco2})
the Hamiltonians are given by
$\Ham\supth{n} 
\equiv \frac{1}{1+2n} (\mapcurv +x\mapcurv_{x})\w\supth{n}
\equiv \frac{1}{1+2n} \Dinvx(\mapcurv\hperp\supth{n})
= \frac{1}{1+2n}\hpar\supth{n}$
(modulo total $x$-derivatives)
for $n\geq 0$,
where $\der{x}\hpar\supth{n}=\mapcurv\hperp\supth{n}$.
Thus $\int\hpar\supth{n} dx$ represents a
(scaled) Hamiltonian functional on the $x$-jet space of $\mapcurv$.
The same expression carries over to the $-1$ flow in the mKdV hierarchy,
$\hperp\supth{-1} = \sin(\theta)$,
$\hpar\supth{-1} = \cos(\theta)$,
given in terms of $\mapcurv=-\theta_{x}$.
Specifically, 
$\hperp\supth{-1} =\Hop(\delta\Ham\supth{-1}/\delta\mapcurv)$
holds for the Hamiltonian 
$\Ham\supth{-1} =-\hpar\supth{-1}$,
with 
$\w\supth{-1} =\delta\Ham\supth{-1}/\delta\mapcurv 
=\Dinvx(\delta\Ham\supth{-1}/\delta\theta)$.

Recent work 
\cite{sigmapaper}
has generalized these results to 
the two known vector versions of the mKdV equation
\cite{SokolovWolf,AncoWolf}, 
by deriving their bi-Hamiltonian integrablity structure
along with their associated hierarchies of 
(higher order) vector soliton equations
from geometric curve flows in the Riemannian symmetric spaces
$SO(N+1)/SO(N)\simeq S^n$ and $SU(N)/SO(N)$.
These two spaces exhaust (cf. \cite{Helgason})
all examples of irreducible symmetric spaces
$G/SO(N)$ given by compact simple Lie groups $G$,
and in the case $N=2$
they both coincide with the classical Riemannian spherical geometry 
$S^2 \simeq SO(3)/SO(2) \simeq SU(2)/SO(2)$.

This geometric derivation of the vector mKdV hierarchies 
involves several main ideas.
Firstly, 
for a non-stretching curve $\map$,
the components of the principal normal 
in an adapted moving frame along $\map$ 
play the role of a natural Hamiltonian flow variable.
Secondly,
the torsion and curvature equations of the Riemannian connections
$\covder{t},\covder{x}$
on the two-dimensional surface of any flow $\map(t,x)$ of such curves
in $G/SO(N)$
geometrically encode Hamiltonian cosymplectic and symplectic operators
$\Hop,\Jop$,
where $x$ is the arclength on $\map$.
In particular, the Hamiltonian structure looks simplest
if the moving frame along $\map$ is chosen 
so that it has a parallel connection matrix \cite{Bishop}
given by an algebraic version of the basic geometric property
$\covder{x} N$ $\parallel$ $T$,
$\covder{x} T$ $\perp$ $T$ 
of an adapted moving frame generalized from $S^2$ to $G/SO(N)$.
For such a moving parallel frame 
the torsion and curvature equations reduce to a simple form
which can be derived directly from the infinitesimal Klein geometry 
of the tangent spaces $\g/\alg{so}(N)$ in $G/SO(N)$, 
where $\g$ is the Lie algebra of $G$.
Thirdly, within the class of non-stretching curves 
whose parallel moving frame 
is preserved by an $O(N-1)$ isotropy subgroup of
the local frame structure group $SO(N)$ of the frame bundle of $G/SO(N)$,
the encoded operators $\Hop,\Eop=\Hop\Jop\Hop$ 
comprise an $O(N-1)$-invariant bi-Hamiltonian pair. 
Moreover, they possess the mKdV scaling symmetry 
and exhibit symmetry invariance under $x$-translations. 
As a result, these operators yield a hereditary recursion operator 
$\Rop=\Hop\Jop$
generating a hierarchy of integrable curve flows in $G/SO(N)$,
organized by their scaling weight,
in which the flow variable satisfies an $O(N-1)$-invariant 
vector evolution equation. 
The $+1$ flow in the respective hierarchies for $G=SO(N+1),SU(N)$
is given by the two known vector mKdV soliton equations \cite{SokolovWolf},
and there is a $-1$ flow described by vector hyperbolic equations
\cite{AncoWolf}
(variants of vector SG equations)
coming from the kernel of the respective mKdV recursion operators
in the two hierarchies.
All the flows in  both hierarchies correspond to 
commuting bi-Hamiltonian vector fields
in the $x$-jet space of the flow variable. 
Finally, 
the evolution equations of the curve $\map(t,x)$ 
produced by the $\pm 1$ flows in each hierarchy 
are $G$-invariant generalizations of the geometric map equations
found for these flows in the case 
$S^2 \simeq SO(3)/SO(2) \simeq SU(2)/SO(2)$,
identified as wave maps and mKdV analogs of Schrodinger maps,
on $G/SO(N)$.

More recently there has been an extension of these results
\cite{imapaper}
deriving complex generalizations of the vector mKdV hierarchies
along with their bi-Hamiltonian integrability structure
from geometric curve flows in the Lie groups
$G=SO(N+1),SU(N)$. 
The derivation adapts the moving parallel frame formulation of 
non-stretching curve flows 
to these Lie groups by viewing them as Riemannian symmetric spaces 
in the standard manner \cite{KobayashiNomizu,Helgason},
with the frame invariance group being 
a natural isotropy subgroup $U(N-1) \subset G$ contained in 
the local structure group of the frame bundle 
of $G=SO(N+1),SU(N)$. 
This leads to bi-Hamiltonian hierarchies of curve flows $\map(t,x)$
in which the flow variable 
(again described by the components of 
the principal normal vector along the curves)
obeys a $U(N-1)$-invariant vector evolution equation.
In particular, 
the hierarchies contain 
the known complex generalizations \cite{SokolovWolf}
of the two versions of vector mKdV soliton equations,
as well as 
complex generalizations of the corresponding vector SG equations
which were first obtained by symmetry-integrability classifications
\cite{AncoWolf}. 

A full generalization of these ideas and results
will be presented here to obtain
bi-Hamiltonian hierarchies of group-invariant soliton equations
arising from geometric curve flows 
in general Riemannian symmetric spaces $G/H$.
All irreducible examples of these spaces divide into two types, 
where either $G$ is a simple Lie group
(and $H$ is a compact subgroup invariant 
under an involutive automorphism of $G$)
or $G$ is a Lie group product $K\times K$
(and $H$ is a diagonal subgroup) such that
$G/H \simeq K$ is a compact simple Lie group.

{\bf Theorem~1.1: }{\it
For each Riemannian symmetric space $G/H$
there is a family of bi-Hamiltonian hierarchies of non-stretching curve flows
described by $G$-invariant geometric maps on $G/H$.
The components of the principal normal vector along these curves
in a moving parallel frame satisfy 
group-invariant (multicomponent) soliton equations of mKdV and SG type,
with an explicit bi-Hamiltonian structure.
In each hierarchy 
the geometric maps corresponding to 
the SG flow and the lowest-order mKdV flow 
are universally given by a non-stretching wave map
and a mKdV analog of a non-stretching Schrodinger map. 
}

{\bf Remark~1.2: }
The invariance group of the soliton equations 
and their bi-Hamiltonian integrability structure 
consists of the linear isotropy subgroup in $\Ad(H)$ that preserves 
the tangent vector for the corresponding curves 
in the tangent space $T_o G/H =\g/\h$ at the origin $o$. 
(By $G$-invariance, any geometric curve flow is equivalent to one that
passes through the origin $o$ in $G/H$.)

{\bf Remark~1.3: }
Up to isomorphism, the distinct bi-Hamiltonian hierarchies in the family
admitted by a given Riemannian symmetric space $G/H$ 
are in one-to-one correspondence with equivalence classes of
unit-norm elements contained in any fixed maximally abelian subspace 
in $\g/\h$ modulo the Weyl group $W \subset \Ad(H)$. 

These results provide a geometric origin and unifying interpretation that
encompasses many examples of multicomponent soliton equations. 
All examples can, moreover, be written down in explicit form,
using Cartan's classification of symmetric spaces \cite{Helgason}.

To begin, some preliminaries are stated in section~2
on the relevant differential geometric and Lie algebraic properties of
symmetric spaces \cite{Helgason,KobayashiNomizu}, 
particularly 
the infinitesimal Klein geometry 
\cite{Sharpe}
attached to $\g/\h$, 
and the Cartan subspaces in $\g/\h$
determined by the Lie algebras $\h,\g$ of $H,G$.
These properties are used in an essential way in section~3
for the study of non-stretching curve flows $\map(t,x)$ in $G/H$,
where $x$ is arclength 
(so $|\mapder{x}|=1$ is the non-stretching property 
using the Riemannian metric on $G/H$).
The main starting point is given by 
the pull back of the torsion and curvature equations of 
the Riemannian connection on $G/H$ 
to the two-dimensional surface of the flow of $\map$,
as derived directly by a frame formulation in terms of 
the infinitesimal Klein geometry of $\g/\h$. 
Based on the decomposition of $\g$ relative to $\ad(e)$
where $e$ is a unit vector belonging to a Cartan subspace in $\g/\h$,
a natural algebraic construction is given for 
a parallel moving frame along $\map$,
evolving under the flow.
Then in section~4, 
the moving frame components of the torsion and curvature 
are shown to encode a bi-Hamiltonian pair of 
cosymplectic and symplectic operators $\Hop,\Jop$
with respect to a Hamiltonian flow variable given by 
the moving frame components of the principal normal along $\map$.
The proof of this main result will involve extending
the standard theory of bi-Hamiltonian structures \cite{Olver,Dorfman}
to the setting of Lie-algebra valued variables. 

The hierarchy of bi-Hamiltonian commuting flows produced by 
the operators $\Hop,\Jop$ is studied in section~5,
where the $+1$ and $-1$ flows are constructed and found to yield
group-invariant multicomponent versions of mKdV and SG equations
associated to all irreducible Riemannian symmetric spaces. 
The geometric map equations produced by these universal $\pm 1$ flows
are derived in detail. 
Concluding remarks on extensions of these results
will be made in section~6.

Some related results of interest have been obtained in recent literature. 
In one direction, 
Hamiltonian operators for one of the vector mKdV equations
as well as for a related scalar-vector KdV system  
have been derived from non-stretching curve flows 
in $N$-dimensional constant curvature Riemannian geometries 
(\ie/ the $N$-sphere)
and in $N$-dimensional flat conformal geometries 
(\ie/ the M\"obius $N$-sphere),
using two different approaches.
One approach 
\cite{SandersWang1,SandersWang2}
viewed these geometries as locally modeled by
Klein geometries on $\Rnum^N$ defined respectively by 
a Euclidean isometry group action $Euc(N)$
and a M\"obius (conformal) isometry group action $Mob(N)\simeq SO(N+1,1)$.
A subsequent derivation 
\cite{Anco1,AncoWang}
used just the intrinsic 
Riemannian and conformal connections defined on the underlying manifold $S^N$. 
Both derivations utilize the approach of studying curve flows via
parallel moving frames, first introduced in 
\cite{LangerPerline,MaribeffaSandersWang,SandersWang1}.
The same approach has also been applied recently \cite{Sanders}
to curve flows in symplectic geometry $Sp(N+1)/Sp(N)\times Sp(1)$,
giving a geometric derivation of 
a quaternionic (non-commutative) mKdV equation.

Earlier fundamental work appeared in \cite{AthorneFordy,Athorne}
on multicomponent mKdV equations derived from symmetric spaces
by algebraic and geometric considerations,
and on investigation of Hamiltonian structures of such equations. 

Another direction 
\cite{ChouQu1,ChouQu2}
has concentrated on deriving the known scalar soliton equations
from non-stretching curve flows in various classical plane geometries:
hyperbolic plane, affine and fully-affine plane, plane similarity geometry.
These geometries together with the Euclidean plane
are characterized by their respective isometry groups
$SL(2)$, $SA(2)$ and $A(2)$, $Sim(2)$, $Euc(2)$,
each acting locally and effectively on $\Rnum^2$;
as such they comprise the main examples of planar Klein geometries
$G/H\simeq \Rnum^2$ 
where $G$ is the isometry group
and $H\subset G$ is a stabilizer subgroup that leaves fixed the origin 
in $\Rnum^2$. 
In this geometric setting
there is a natural group-invariant notion of geometric curve flows
based on the evolution of differential invariants 
\cite{Olver}
of a non-stretching curve $\map$,
formulated using a moving frame $(T,N)=(\mapder{x},\mapder{xx})$
where $x$ is a natural $G$-invariant arclength on $\map$.
Generalizations to non-stretching flows of space curves 
in Klein geometries on $\Rnum^3$
have also been studied in a few cases \cite{ChouQu3,ChouQu4}:
Euclidean space $Euc(3)$, similarity geometry $Sim(3)$,
affine space $A(3)$, centro-affine space $SL(3)$. 
The results obtained to-date have provided an elegant geometric origin
for many scalar soliton equations. 
For instance, 
the KdV equation and the Sawada-Kotera equation 
are noted to arise respectively as 
a centro-affine and affine version of the mKdV equation 
coming from Euclidean geometry. 
It should be possible to derive the bi-Hamiltonian integrability structure of
these solitons equations by adapting the parallel moving frame approach
used in the present work to the setting of affine geometry 
and similarity geometry.

\section{Preliminaries}

A symmetric space $M=G/H$ is {\it irreducible} if 
it is not a product of smaller symmetric spaces. 
All irreducible Riemannian symmetric spaces have a classification into
two types \cite{KobayashiNomizu,Helgason}:

(I) 
$G$ is a connected simple Lie group
and $H\subset G$ is a compact Lie subgroup
invariant under an involutive automorphism $\sigma$ of $G$
(in this case $M$ is compact or noncompact according to whether $G$ is).
\vskip0pt
(II)
$G$ is a product $K\times K$ for a connected compact simple Lie group $K$
and $H$ is its diagonal product subgroup in $G$ on which 
a permutation of the product factors 
acts as an involutive automorphism $\sigma$,
with $G/H \simeq K \simeq H$
(in this case, since $G$ is compact, so is $M$).

Henceforth let $M=G/H$ be an irreducible Riemannian symmetric space
in which $G$ acts effectively on the manifold $M$ by left multiplication.
The automorphism $\sigma$ induces a decomposition on 
the Lie algebra of $G$, given by a direct sum of vector spaces
$\g = \h \oplus \m$ 
such that $\sigma(\h) =\h$, $\sigma(\m) =-\m$,
with Lie bracket relations 
$[\h,\h]_\g \subset \h$,
$[\h,\m]_\g \subset \alg{\m}$,
$[\m,\m]_\g \subset \alg{\h}$,
where $\h$ is identified with the Lie algebra of $H$
and $\m =T_\o M$ is identified with the tangent space 
at the origin $\o$ in $M$.
The group action of $G$ on $M$ provides a canonical isomorphism 
$T_x M \simeq_G \m$.

\subsection{Riemannian metric and connection}

The Riemannian structure of $G/H$ comes from the Cartan-Killing form 
$\langle \cdot,\cdot\rangle $ of $\g$.
Since $\g$ is a simple Lie algebra,
the subspaces $\h,\m$ are orthogonal $\langle \h,\m\rangle =0$ 
and hence the Cartan-Killing form restricts to yield 
a nondegenerate inner product on $\m$
which is $\Ad_\g(H)$ invariant. 
This inner product $\langle \cdot,\cdot\rangle_\m$ 
has a definite sign, negative or positive,
according to whether $\g$ is compact or noncompact.
In the noncompact case, 
the Lie algebra $\g=\h\oplus\m$ is dual to $\g^*=\h\oplus\i\m$
which defines a compact real Lie algebra 
contained in the complexification of $\g$.
This leads to a corresponding duality \cite{KobayashiNomizu,Helgason}
between compact and noncompact
Riemannian symmetric spaces of type I. 
As there are no noncompact Riemannian symmetric spaces of type II,
we will henceforth restrict attention to the compact case
in order to include both types I and II. 

Under the canonical identification $\m \simeq_G T_x M$,
the negative-definite inner product $\langle \cdot,\cdot\rangle_\m$
defines a Riemannian metric tensor $g$ on $M$ given by 
\EQ
g(X,Y) = -\langle X,Y\rangle_\m
\endEQ
for all $G$-invariant vector fields $X,Y$ in $T_x M$.
Associated to this metric $g$ is a unique torsion-free $G$-invariant
Riemannian connection $\covder{}$,
determined by $\covder{}g=0$,
where 
\EQ
\tors(X,Y)= \covder{X}Y-\covder{Y}X -[X,Y]=0
\endEQ
is the torsion-free property.
Here $[\cdot,\cdot]$ denotes the commutator of vector fields on $M$.
The Riemann curvature tensor of the connection $\covder{}$ is given by 
\EQ
R(X,Y) = [\covder{X},\covder{Y}] -\covder{[X,Y]}
= -\ad_\m([X,Y]_\m)
\endEQ
with $\ad_\m$ viewed as a linear map on $T_x M \simeq_G \m$.
This curvature tensor is $G$-invariant 
and covariantly constant $\covder{}R=0$.
Note that left multiplication by $G$ acting on $M$
generates the isometries of the manifold $M$,
namely the isometry group of $g$ is $G$.

Recall, the linear isotropy group $H^*$ of $G/H$ is
the representation of $H$ by linear transformations on 
$T_\o M$ at the origin $\o$ in $M$. 
This group is isomorphic to $\Ad_\m(H)$
through the identification $T_\o M =\m$.

\subsection{Klein geometry and soldering frames}

For any Riemannian symmetric space $M=G/H$, 
the Lie group $G$ has the structure of a principal $H$-bundle over $M$,
$\pi:G \rightarrow G/H$ \cite{KobayashiNomizu,Sharpe}. 
Consider the Maurer-Cartan form $w_G$ on $G$,
which is a left-invariant $\g$-valued $1$-form 
that provides the canonical identification $T_\grpid G =\g$
where $\grpid$ is the identity element of $G$.
The Maurer-Cartan form satisfies 
$\d w_G + \frac{1}{2}[w_G,w_G]_\g =0$
and hence, viewed geometrically, it is a zero-curvature connection 
on the bundle $G$.
This collective structure is called \cite{Sharpe} 
the Klein geometry of $M=G/H$.
It has a useful reformulation locally on $M$
involving the Lie algebra structure of the tangent spaces of $G/H$.

Let $\psi:U \rightarrow \pi^{-1}(U) \subset G$ 
be any local section of this bundle, where $(U,\psi)$ are coordinates
for a local trivialization 
$G \approx U\times H$.
The tangent space of the trivialization $U\times H$ at any point $(x,\psi(x))$
is isomorphic to the Lie algebra $\g$ of $G$.
In particular, note 
$\pi^*:\g \simeq \m\times\h \rightarrow \m$ as vector spaces,
with 
$T_x M \simeq_G \m =\g/\h$, 
$T_{\psi(x)} H \simeq \h$
and $T_{(x,\psi(x))} G \simeq \g =\m\oplus\h$.
Now, the pullback of the Maurer-Cartan form $w_G$ by $\psi$
yields a $\g$-valued $1$-form $\gconx$ (locally) on $M$, 
satisfying the Cartan structure equation
\EQ\label{zerocurveq}
D \gconx +\frac{1}{2}[\gconx,\gconx]_\g =0
\endEQ
where $D$ is the total exterior derivative operator on $M$
and $[\cdot,\cdot]_\g$ is understood to denote 
the Lie algebra bracket in $\g$ composed with the wedge product in $T^*_x M$.
(Informally, 
we can view $\gconx :T_x M \simeq_G \m =\g/\h \rightarrow \g$
as attaching an infinitesimal Klein geometry to $\g/\h$.)
Note a change of the local section $\psi$,
described by a smooth function $h:U \rightarrow H$ 
yielding the local section  
$\tilde\psi = \psi h$,
induces a change of $\gconx$ given by 
\EQ
\widetilde\gconx =\Ad(h^{-1})\gconx +h^{-1} Dh . 
\endEQ
This is called \cite{Sharpe} a gauge transformation on $\gconx$
relative to the gauge $\psi$.

Introduce 
\EQ
\e := \frac{1}{2}(\gconx -\sigma(\gconx)) ,\qquad
\conx := \frac{1}{2}(\gconx +\sigma(\gconx)) ,
\endEQ
denoting the even and odd parts of $\gconx$,
defined using the automorphism $\sigma$ of $\g$.
The corresponding decomposition under a gauge transformation on $\gconx$
yields
\EQ\label{gaugetransf}
\tilde\e =\Ad(h^{-1})\e ,\qquad
\tilde\conx = \Ad(h^{-1})\conx +h^{-1} Dh . 
\endEQ

{\bf Proposition~2.1: }{\it
On $U\subset M$,
the $\m$-valued $1$-form $\e$ represents a linear coframe
and the $\h$-valued $1$-form $\conx$ represents a linear connection,
with $H$ as the structure (gauge) group defining a bundle of linear frames.
}

Let the $2$-forms 
\EQ
\frtors=D\e+[\conx,\e]_\g =\frder\e 
\quad\text{ and }\quad
\frcurv=D\conx+\frac{1}{2} [\conx,\conx]_\h =[\frder,\frder]
\endEQ
denote the $\m$-valued torsion and $\h$-valued curvature
associated to the linear connection and coframe,
where 
\EQ\label{frcovder}
\frder = D +[\conx,\cdot]_\g
\endEQ
is the (gauge) covariant exterior derivative.

{\bf Proposition~2.2: }{\it
The Cartan structure equation \eqref{zerocurveq}
decomposes with respect to $\sigma$ to give
\EQ\label{cartaneqs}
\frtors=0 ,\quad 
\frcurv=-\frac{1}{2}[\e,\e]_\m ,
\endEQ
so thus $\frder$ is torsion-free and has covariantly-constant curvature. 
}

The precise interpretation of $\e$ and $\conx$ in these propositions
comes from a basis expansion 
$\e =\e_{a}\mvec{a}$,
$\conx =\conx_{i}\hvec{i}$,
where $\mvec{a},\hvec{i}$ are any fixed orthonormal basis for $\m,\h$
with respect to the Cartan-Killing inner product,
\EQ\label{onbasis}
\langle\mvec{a},\mvec{b}\rangle =-\delta^{ab} ,\quad
\langle\hvec{i},\hvec{j}\rangle =-\delta^{ij} ,\quad
\langle\mvec{a},\hvec{i}\rangle =0 .
\endEQ
Then $\{\e_{a}\}$ will define frame covectors spanning
the cotangent space of $M$ at $x$,
and $\{\conxhook{i}\}$ will define connection $1$-forms related to 
the exterior derivative of the frame covectors 
by the torsion-free property of the covariant derivative $\frder$ 
on $M$ at $x$.

Most importantly, 
$\e$ and $\conx$ provide a natural soldering of
this frame geometry of $M=G/H$ locally onto 
the Riemannian geometry of $M=G/H$.
Let $\duale$ be dual to $\e$,
namely a $\m$-valued vector such that 
$-\langle \duale,\e\rangle_\m =\id_x$ is the identity map on $T_x M$.
Then $\duale$ represents an orthonormal linear frame on $U\subset M$,
so thus $\{\duale_{a}\}$ 
as given by a basis expansion 
$\duale =\duale_{a}\mvec{a}$
will define frame vectors spanning the tangent space of $M$ at $x$
and obeying the orthonormality property 
$\duale_{a}\hook\e_{b} = \delta_{ab}$.

{\bf Theorem~2.3: }{\it
The linear coframe $\e$ and linear connection $\conx$
determine a Riemannian metric and Riemannian connection on $M=G/H$
given by 
\EQ\label{solder}
g=-\langle \e,\e\rangle_\m ,\quad
\covder{}\cdot = -\langle \duale,\frder(\e\hook\cdot)\rangle_\m
= -\langle \duale,D(\e\hook\cdot)+[\conx,(\e\hook\cdot)]_\g\rangle_\m .
\endEQ
In particular, for all $X,Y$ in $T_x M$,
\EQ
g(X,Y) = -\langle \ehook{X},\ehook{Y}\rangle_\m 
\endEQ
and 
\EQ
\covder{X}Y = -\langle \duale,\frderhook{X}\ehook{Y} \rangle_\m ,
\quad\text{or equivalently}\quad
\frderhook{X}\ehook{Y} = \e\hook\covder{X}Y
\endEQ
where $\ehook{X} =\e\hook X,\ehook{Y} =\e\hook Y$ 
are the coframe projections of $X,Y$. 
Moreover, the Riemannian curvature is determined by the curvature of
the exterior covariant derivative $\frder$,
\EQ
\riem(X,Y)Z 
= -\langle \duale,[\frderhook{X},\frderhook{Y}]\ehook{Z}-\frderhook{[X,Y]}\ehook{Z}\rangle_\m
= -\langle \duale,[\frcurvhook{X,Y},\ehook{Z}]_\g\rangle_\m
\endEQ
with (cf Proposition~2.2)
\EQ
\frcurvhook{X,Y} =\frcurv\hook(X\wedge Y) = -[\ehook{X},\ehook{Y}]_\m .
\endEQ
Hence 
\EQ
\riem(X,Y) 
= \langle [[\ehook{X},\ehook{Y}]_\m,\e]_\g,\duale\rangle_\m ,
\endEQ
or equivalently
\EQ
\riem(\cdot,\cdot) 
= -\langle [\e\hook\cdot,\e\hook\cdot]_\m,[\duale,\e]_\m\rangle_\h . 
\endEQ
}

Here note $[\duale,\e]_\m$ defines a $\h$-valued linear map on $T_x M$
displaying skew symmetry with respect to $g$:
$g(Y,[\duale,\e]_\m X) = [\ehook{Y},\ehook{X}] = -g(X,[\duale,\e]_\m Y)$.

{\it Remark: }
The torsion-free property of the Riemannian connection is expressed by
(cf Proposition~2.2)
\EQ
\tors(X,Y) 
= -\langle \duale,(\frderhook{Y}\e)\hook X - (\frderhook{X}\e)\hook Y\rangle_\m
= -\langle \duale,\frtors_{X,Y}\rangle_\m
=0
\endEQ
with 
\EQ
\frtorshook{X,Y} = \frtors\hook(X\wedge Y) . 
\endEQ

The soldering structure of the Riemannian connection has
the following alternative formulation.

{\bf Corollary~2.4: }{\it
For all $X,Y$ in $T_x M$,
\EQ
\covder{X}\e =[\e,\conxhook{X}]_\g ,\quad
[\covder{X},\covder{Y}]\e -\covder{[X,Y]}\e=-\e\hook\riem(X,Y)
=[\e,\frcurvhook{X,Y}]_\g
\endEQ
where $\conxhook{X} =X\hook\conx$ is the linear connection 
projected in the $X$ direction on $M$. 
}

In examples, it will be convenient to work in a matrix representation
for $\m,\h,\g$ viewed as subspaces in $\alg{gl}(N,\Rnum)$
for suitable $N \geq 1$. 
Hence $\mvec{a},\hvec{i}$ will then represent $N\times N$ matrices
satisfying the commutator relations
\EQ\label{commbasis}
[\hvec{i},\hvec{j}] = \c{ij}{k}\hvec{k} ,\quad
[\mvec{a},\mvec{b}] = \c{ab}{i}\hvec{i} ,\quad
[\hvec{i},\mvec{a}] = \c{ia}{b}\mvec{b} ,
\endEQ
where $\c{ij}{k},\c{ab}{i}\c{ia}{b}$ are the structure constants of $\g$
with respect to the matrix basis $\{\mvec{a},\hvec{i}\}$.
More directly, $\h$ and $\g$ can be represented as matrix subalgebras
in $\alg{gl}(N,\Rnum)$, with $\m$ represented as the matrix subspace $\g/\h$
in $\alg{gl}(N,\Rnum)$. 
Then the soldering structure is given by matrix formulas
\EQs
&& 
g(X,Y) = -\chi\tr(\ehook{X}\ehook{Y}) ,
\\
&&
\covder{X}Y = -\chi\tr(\duale\frderhook{X}\ehook{Y}) ,\quad
\riem(X,Y) = -\chi\tr([\duale,\e][\ehook{X},\ehook{Y}]) ,
\endEQs
for some constant $\chi$ depending on 
the choice of matrix representation of $\m$,
where 
$\ehook{X} = (\e_a\hook X)\mvec{a}$, $\ehook{Y} = (\e_a\hook Y)\mvec{a}$
are $N\times N$ matrices belonging to $\m$; 
$\e=\e_a \mvec{a}$, $\duale=\duale_a \mvec{a}$
are $N\times N$ vector-valued matrices belonging to 
$T_x^* M\otimes \m$, $T_x M\otimes \m$ respectively. 
Moreover, 
\EQs
&& 
\covder{X}\duale_a = -\c{ib}{a} (\conxhook{X})_i \duale_b , 
\label{soldere}\\
&& 
\riem(X,Y)\duale_a = -\c{ib}{a} (\frcurvhook{X,Y})_i \duale_b 
= \c{cd}{i}\c{ib}{a} X_c Y_d \duale_b , 
\endEQs
with $X_a = \e_a\hook X,Y_a = \e_a\hook Y$. 

Due to the soldering, the frame covectors obey the orthonormality property
$g(\duale_{a},\duale_{b})^{-1}= -\tr(\mvec{a}\mvec{b})$.

\subsection{Action of $\bf ad(\it e)$}

It is useful to consider an algebraic decomposition of $\g$
with respect to $\ad(\e)$ where $\e$ is the soldering frame. 
Let $X$ be a vector in $T_x M$ and consider the element 
$\ehook{X} =\e\hook X$ given by the frame projection of $X$
into $\m \subset \g$. 
The centralizer of $\ehook{X}$ is the set $\cen(\ehook{X})$ of 
all elements annihilated by $\ad(\ehook{X})$
in the Lie algebra $\g$, namely $[\ehook{X},\cen(\ehook{X})]=0$. 
This set has the obvious structure of a Lie subalgebra of $\g$,
since $\ad(\ehook{X})[\cen(\ehook{X}),\cen(\ehook{X})]=0$
from the fact that $\ad$ acts as a derivation on the Lie bracket of $\g$. 
Write $\mcen,\hcen$ for the intersection of $\cen(\ehook{X})$ 
with the subspaces $\m,\h$ in $\g=\m\oplus\h$,
and $\mcenperp,\hcenperp$ for their respective orthogonal complements
(perp spaces) where
\EQ
\m=\mcen\oplus\mcenperp ,\qquad
\h=\hcen\oplus\hcenperp . 
\endEQ
Thus, note 
$[\ehook{X},\mcen]=[\ehook{X},\hcen]=0$,
while 
$[\ehook{X},\mcenperp]\neq 0$, $[\ehook{X},\hcen]\neq 0$.

{\bf Proposition~2.5: }{\it 
The centralizer subspaces $\mcen,\hcen$ of $\ehook{X}$
have the Lie bracket relations
\EQs
&&
[\hcen,\hcen] \subseteq \hcen ,
\\&&
[\hcen,\mcen] \subseteq \mcen ,\quad
[\mcen,\mcen] \subseteq \hcen .
\endEQs
Hence the perp spaces satisfy the relations
\EQs
&&
[\hcen,\hcenperp] \subseteq \hcenperp ,\quad
[\mcen,\mcenperp] \subseteq \hcenperp ,
\label{perpspbracketh}\\
&&
[\hcen,\mcenperp] \subseteq \mcenperp ,\quad
[\mcen,\hcenperp] \subseteq \mcenperp .
\label{perpspbracketm}
\endEQs
}

The proof uses just the Lie bracket structure of $\g$
along with the invariance of the Cartan-Killing inner product.
For instance, consider $[\hcen,\hcen]$. 
Because $\ad(\ehook{X})$ is a derivation it annihilates $[\hcen,\hcen]$
and hence, since $\h$ is a Lie subalgebra of $\g$, 
$[\hcen,\hcen]$ is contained in $\hcen$.
Thus $[\hcen,\hcen]$ is orthogonal to $\hcenperp$, 
and so by invariance, 
$0= \langle \hcenperp,[\hcen,\hcen]\rangle_\h = \langle \hcen,[\hcen,\hcenperp]\rangle_\h$
which implies $[\hcen,\hcenperp]$ is contained in $\hcenperp$.
The other relations are established similarly. 

The observation that $\ehook{X}$ lies in $\mcen$,
combined with the Lie bracket relations in Proposition~2.5,
shows that $\ad(\ehook{X})$ maps $\hcen$ into $\mcen$, 
$\hcenperp$ into $\mcenperp$, and vice versa. 
Consequently, $\ad(\ehook{X})^2$ is well-defined as a linear mapping of
the subspaces $\hcenperp$ and $\mcenperp$ into themselves. 
These subspaces are likewise taken into themselves by 
all the linear maps $\ad(\hcen)$, corresponding to 
the infinitesimal action of the group of linear transformations 
that preserves $\ehook{X}$. 
In particular,
let $\isoH$ be the closed subgroup in the linear isotropy group $H^*$ 
given by $\Ad(h) \ehook{X} = \ehook{X}$. 
Since $[\ad(\ehook{X}),\ad(\hcen)]=\ad([\ehook{X},\hcen]=0$,
note that $\hcen$ is isomorphic to the Lie algebra of $\isoH$
and that the linear transformations in this group
commute with the linear map $\ad(\ehook{X})^2$. 
Schur's lemma applied to $\ad(\ehook{X})^2$ now gives the following result. 

{\bf Proposition~2.6: }{\it 
(i) 
$\mcenperp$ and $\hcenperp$ are isomorphic as vector spaces
under the linear map $\ad(\ehook{X})$. 
(ii) 
$\mcenperp$ and $\hcenperp$ each decompose into 
a direct sum of subspaces given by irreducible representations of 
the group $\isoH$ 
on which the linear map $\ad(\ehook{X})^2$ is a multiple of the identity.
}

Note $\dim\mcenperp = \dim\hcenperp =\dim \m -\dim \isoH$,
so these spaces will be of minimal dimension 
(depending on $\ehook{X}$) 
precisely when the group $\isoH$ is of maximal size. 

The decomposition of these vector spaces in Proposition~2.6
will be maximally irreducible whenever the linear map 
$\ad(\ehook{X})^2$
has equal eigenvalues, whereby $\ad(\ehook{X})^2=\chi\id_\perp$
is a multiple of the identity map $\id_\perp$ on 
$\mcenperp \simeq\hcenperp$. 
In this situation, additional Lie bracket relations can be established
as follows. 
Introduce the bracket $(\cdot,\cdot):=[\ad(\ehook{X})\cdot\ ,\cdot\ ]$
on the centralizer subspace 
$\gcenperp:=\mcenperp\oplus\hcenperp$ of $\ehook{X}$ in $\g$, 
and write $I_X:= \ad(\ehook{X})^2$. 
Since $\ad(\ehook{X})$ acts as a derivation, note
\EQs
I_X(y,z)
&=& 
[\ad(\ehook{X})I_X y,z]
+ [\ad(\ehook{X})y, I_X z]
+ 2[I_X y, \ad(\ehook{X})z]
\nonumber\\
&=&
(I_X y,z) + (y,I_X z) -2(z,I_X y)
\label{adsqid}
\endEQs
for all vectors $y,z$ in $\gcenperp$. 
Now suppose $I_X=\chi\id_\perp$,
where $\chi$ is a non-zero constant. 
Then the identity \eqref{adsqid} yields
$\chi\id_\perp(y,z)=2\chi(y,z)-2\chi(z,y)$
and hence
$(y,z)_\perp = 2(z,y)_\perp$
where a subscript $\perp$ denotes projection onto $\gcenperp$. 
This identity implies $(y,y)_\perp=0$, 
and hence by polarization, 
$(y,z)_\perp=0$.
As a result, it follows that 
$0=(y,z)_\perp- (z,y)_\perp 
= [\ad(\ehook{X})y,z]_\perp+ [y,\ad(\ehook{X})z]_\perp
= \id_\perp\ad(\ehook{X})[y,z]$
whence
$[y,z]_\perp=0$
since $\ad(\ehook{X})$ is nondegenerate on $\gcenperp$.

{\bf Proposition~2.7: }{\it
If $\ad(\ehook{X})^2=\chi\id_\perp$ is multiple of the identity 
on $\mcenperp \simeq\hcenperp$, then these centralizer subspaces obey
the Lie bracket relations
\EQs
&&
[\hcenperp,\hcenperp] \subseteq \hcen ,\quad
[\mcenperp,\mcenperp] \subseteq \hcen ,
\\&&
[\mcenperp,\hcenperp] \subseteq \mcen .
\endEQs
}

\subsection{Cartan spaces}

The algebraic structure of the spaces $\hcenperp$ and $\mcenperp$ 
along with the group $\isoH$ 
will have a key role in the later construction of 
a parallel moving frame formulation for curve flows
in Riemannian symmetric spaces. 
Part of this algebraic structure is effectively independent of $\ehook{X}$.

Let $\cen_0(\ehook{X})$ denote the center of $\cen(\ehook{X})$,
which is an abelian subalgebra of $\g$.
Write $\cen_0(\ehook{X})_\m$, $\cen_0(\ehook{X})_\h$, 
for the intersections of $\cen_0(\ehook{X})$ with $\m,\h$,
and write $\a$ for any maximal abelian subspace in $\m$. 

It can be shown \cite{Helgason} that 
the image of any maximal abelian subspace $\a$ under the group $H^*$
is the vector space $\Ad(H)\a =\m$.
Hence $\ehook{X}$ is contained in some such subspace,
namely $\a = \cen_0(\ehook{X})_\m \subseteq \mcen \subset \m$.
These subspaces $\a \subset \m$ are known to have 
the following characterization \cite{Helgason}. 

{\bf Lemma~2.8: }{\it
(i) 
Any two maximal abelian subspaces $\a$ are isomorphic to one another
under some linear transformation in $H^* \simeq \Ad(H)$.
(ii) 
Every maximal abelian subspace $\a$ is invariant under the Weyl group 
$W=C(\a)/N(\a)$, which is a finite subgroup contained in
the linear isotropy group $H^*$,
where $C(\a)$ and $N(\a)$ are respectively 
the centralizer group and normalizer group of $\a$ in $H^*$. 
(iii)
Any maximal abelian subspace $\a$ 
is the $-1$ eigenspace in some Cartan subalgebra of the Lie algebra 
$\g =\m\oplus\h$ under the grading $\sigma(\m)=-\m,\sigma(\h)=\h$,
and hence $\a$ is the centralizer of its elements in $\m$, 
$\a =\cen(\a) \subset \m$. 
}

Henceforth the subspaces $\a\subset \m$ will be referred to as
{\it Cartan subspaces} of $\m$ in $\g$.
The close relationship between the subspaces $\a$
and the Cartan subalgebras of $\g$ 
leads to a more explicit description of 
the algebraic decomposition of $\g$ with respect to $\ad(\ehook{X})^2$. 

Let $\g^\Cnum$ be the complexification of $\g$.
Recall, a root of $\g$ is a linear function, $\alpha$, 
from any Cartan subalgebra $\g^0$ of $\g^\Cnum$ into $\Cnum$,
such that for all vectors $w$ in $\g^0$, 
$\ad(w)$ has an eigenvector with the eigenvalue $\alpha(w)$. 
Let the set of all roots of $\g$ be denoted $\Delta$.
The following lemma is adapted from results in \cite{Helgason},
utilizing the fact that $\mcenperp$ is contained in $\a^\perp$
as a consequence of the inclusion $\a \subseteq \mcen$. 

{\bf Lemma~2.9: }{\it
Let $\Delta_\m$ be the set of roots $\alpha$ of $\g$
such that $\alpha(a) \neq 0$ for some $a$ in a Cartan subspace 
$\a \subset \m$.
Since $\g$ is compact, $\alpha(a)$ will be an imaginary function
(namely $\alpha : \a \rightarrow \i \Rnum$). 
Then there exists a basis of vectors 
$\{y_\alpha\}$ in $\mcenperp$ and $\{z_\alpha\}$ in $\hcenperp$,
indexed by a certain subset of the roots $\alpha$ in $\Delta_\m$, 
satisfying the Lie bracket relations
\EQ
[y_\alpha,a]_\g = \i\alpha(a) z_\alpha ,\quad
[z_\alpha,a]_\g = -\i\alpha(a) y_\alpha 
\endEQ
for all vectors $a$ in $\a$.
Hence the eigenvalues of the linear map $\ad(a)^2$ 
on its eigenspaces in $\mcenperp,\hcenperp$ 
are given by $\alpha(a)^2 < 0$
(cf Proposition~2.6). 
Note $\ad(a)^2$ will be a multiple of the identity precisely 
when its eigenvalues $\alpha(a)^2$ are equal. 
}

\section{Parallel moving frames and bi-Hamiltonian operators}

On  a $n$-dimensional Riemannian manifold $M$,
a frame is a basis of the tangent space $T_x M$
at each point $x$ in $M$,
and a coframe is a dual basis of $T^*_x M$.
Let $\{\e_a\}$ be an orthonormal coframe 
and $\{\duale_a\}$ its dual frame, $a=1,\ldots,n$,
namely 
$g(\e_a,\e_b) =\duale_{a}\hook\e_{b} = \delta_{ab}$.
Covariant derivatives of the frame in any direction $X$ at a point in $M$
are given in terms of the frame connection $1$-forms
$\conx_a{}^b = -\conx_b{}^a$ by 
$\covder{X} \e_a = -(\conx_a{}^b \hook X) e_b$. 
Here the frame structure group (modulo reflections) is 
the rotation group $SO(n)$. 
A moving frame along a curve $\map(x)$ in $M$ is
the restriction of a frame to $T_\map M$,
or correspondingly a coframe to $T^*_\map M$, 
describing a cross section of the orthonormal frame bundle over $\map$.
$SO(n)$ gauge transformations allow a moving frame 
to be adapted to $\map$ such that 
\EQ
\duale_a = 
\begin{cases}
\duale_\parallel ,& a=1; \cr
(\duale_\perp)_a ,& a=2,\ldots,n
\end{cases}
\endEQ
where 
$\duale_\parallel=X$ is the unit tangent vector along $\map$
and each $(\duale_\perp)_a$ is a unit normal vector with respect to $\map$.
An adapted frame is preserved by the $SO(n-1)$ group of local rotations
in the normal space of $\map$.
This gauge freedom can be used to geometrically adapt
the connection matrix 
\EQ
(\conxhook{X})_{ab} := \conx_{ab}\hook X
= g(\duale_a,\covder{x} \duale_b)
\endEQ
to $\map$ such that 
the derivative of the tangent vector in the frame is normal to $\map$
while the derivative of each normal vector is tangential to $\map$.
Such a moving frame is said to be {\it parallel}\cite{Bishop}:
\EQ\label{parallelprop}
\covder{x} \duale_\parallel\ \perp\ \duale_\parallel ,\quad
\covder{x} (\duale_\perp)_a\ \parallel\ \duale_\parallel .
\endEQ
Note the connection matrix $\conxhook{X}^{ab}$ of a parallel moving frame
is skew-symmetric with only its top row and left column being nonzero,
as a consequence of the parallel property
\EQ\label{riemparallel}
g(\duale_\parallel,\covder{x}\duale_\parallel) =0 ,\quad
g((\duale_\perp)_a,\covder{x}(\duale_\perp)_b) =0 . 
\endEQ
This property is preserved by any rigid ($x$-independent) 
$SO(n-1)$ rotation in the normal space of $\map$,
which thereby defines the equivalence group for parallel moving frames. 

Any parallel moving frame differs from a classical Frenet frame 
\cite{Guggenheimer}, if $n>2$,
by an $x$-dependent $SO(n-1)$ rotation 
acting in the normal space of the curve $\map$
(see \cite{Wang} for a geometric interpretation of this rotation). 
It is well known that the components of a Frenet connection matrix
along $\map$ are differential invariants of the curve 
\cite{Guggenheimer}. 
Consequently, the components of a parallel connection matrix are thus 
differential covariants of $\map$
in sense that they are invariantly defined up to the covariant action of 
the equivalence group of rigid $SO(n-1)$ rotations. 

There is a purely algebraic characterization of 
the connection matrix of a parallel moving frame 
\cite{sigmapaper}. 
Fix the standard orthonormal basis of $\Rnum^n$,
given by $n$ row vectors $\mvec{a}$ 
whose $a\th$ component is $1$,
with all other components equal to $0$,
where $a=1,\ldots,n$.
Similarly, fix the standard basis of the Lie algebra $\alg{so}(n)$
induced from the isomorphism 
$\Rnum^n \wedge \Rnum^n \simeq \alg{so}(n)$,
where a wedge denotes the vector outer product. 
(Note the natural action of $\alg{so}(n)$ on $\Rnum^n$ is defined here 
by right multiplication of skew matrices on row vectors.)
This basis consists of $n(n-1)/2$ skew matrices 
$\hvec{i}= \mvec{a}\wedge\mvec{b}$
whose only nonzero components are 
$1$ in the $a\th$ row and $b\th$ column,
and $-1$ in the $b\th$ row and $a\th$ column,
where $i$ is identified with $(a,b)$ such that $a<b$, 
for $a,b=1,\ldots,n$.

Now, there are associated to the frame $\duale_a$ 
and its covariant derivatives
the $\Rnum^n$-valued linear coframe 
$\e=\e_a \mvec{a}$
and the $\alg{so}(n)$-valued linear connection $1$-form 
$\conx=\conx_{ab} \mvec{a}\wedge\mvec{b}$ 
satisfying soldering relations \cite{KobayashiNomizu}
analogous to the ones \eqref{solder},
where $\e$ represents a linear map from $\Rnum^n$ onto $T_x M$.
These relations state that $\covder{}$ 
has no torsion
\EQ
\d\e -\e\wedge\conx = \frtors =0
\endEQ
and carries curvature
\EQ
\d\conx +[\conx,\conx]_{\alg{so}(n)} =\frcurv
\endEQ
determined by the frame components of 
the Riemann curvature tensor $[\covder{},\covder{}]$ of the metric $g$ on $M$.

Write 
$\mvec{}_\parallel :=\mvec{1}$ when $a=1$, 
$\mvec{a}_\perp :=\mvec{a}$ when $a=2,\ldots,n$,
so thus 
$\e_\parallel = \e_1 \mvec{}_\parallel$
and
$\e_\perp = \e_a \mvec{a}_\perp$
describe the tangential and normal parts of the linear coframe 
$\e =\e_\parallel +\e_\perp$
relative to $X$,
with $\e_\perp\hook X =0$ for the tangent vector $X$
and $\e_\parallel\hook Y =0$ for all normal vectors $Y$
along $\map$. 
Consider the derivatives of the coframe $\e$ along $\map$
and decompose the connection matrix 
$\conxhook{X} := \conx\hook{X} = \conxhook{X}^0 + \conxhook{X}^\perp$
into separate parts given by 
$\conxhook{X}^0 
= (\conxhook{X})_{ab} \mvec{a}_\perp\wedge\mvec{b}_\perp
=0$
and 
$\conxhook{X}^\perp 
= (\conxhook{X})_{1a} \mvec{}_\parallel\wedge\mvec{a}_\perp
\neq 0$.
Observe that in this decomposition 
$\hvec{i}_\perp :=\mvec{a}_\perp\wedge\mvec{b}_\perp$
is a basis for the Lie subalgebra 
$\alg{so}(n-1) \simeq \Rnum^{n-1}\wedge\Rnum^{n-1}$
where $\Rnum^{n-1} \subset \Rnum^n$ is 
the subspace orthogonal to $\mvec{}_\parallel$,
while 
$\hvec{i}_\parallel := \mvec{}_\parallel\wedge\mvec{a}_\perp$
is a basis for the perp space of $\alg{so}(n-1) \subset \alg{so}(n)$. 

{\bf Proposition~3.1: }{\it
A linear coframe $\e =\e_\parallel +\e_\perp$ along a curve $\map(x)$
in a Riemannian manifold $M$ is a parallel moving frame iff
its connection matrix $\conxhook{X}$ vanishes on 
the Lie subalgebra $\alg{so}(n-1) \subset \alg{so}(n)$ of
the $SO(n-1)$ rotation subgroup that leaves the tangent covector $\e_\parallel$
invariant under the $SO(n)$ frame structure group. 
}

This algebraic characterization of the connection matrix of 
a moving parallel frame for curves in Riemannian geometry 
has a direct generalization to the setting of Klein geometry 
following the ideas in \cite{sigmapaper}.

\subsection{Construction of parallel moving frames in Klein geometry}

Let $\map(x)$ be a smooth curve 
in some $n$-dimensional compact Riemannian symmetric space $M=G/H$,
and write $X=\mapder{x}$ for its tangent vector 
with an affine parametrization so that $x$ is the arclength. 
Locally on $M$, introduce 
a $\m$-valued linear coframe $\e$ 
and a $\h$-valued linear connection $\conx$
satisfying the soldering relations \eqref{solder}
in terms of the Riemannian metric $g$ and connection $\covder{}$
on the manifold $M$,
where $\m=\g/\h$ is the vector space quotient of the Lie algebras $\g,\h$. 
This structure naturally arises from 
the Klein geometry of the Lie group $G$
viewed as a principle bundle over $M$ whose structure group is
the Lie subgroup $H \subset G$. 
The projections of $\e$ and $\conx$ along the curve $\map(x)$
\EQ
\ehook{X} := \e\hook X ,\qquad
\conxhook{X} := \conx\hook X 
\endEQ
provide a moving frame formulation of this curve
such that the frame structure group is
the linear isotropy group $H^* \simeq \Ad(H)$
acting as local gauge transformations \eqref{gaugetransf}. 
The arclength property of $x$ in this formulation is expressed by 
$|X|_g =|\ehook{X}|_\m=1$ 
so $\ehook{X}$ has unit norm in the Cartan-Killing metric on $\m$
(where $|\ehook{X}|^2_\m=-\langle \ehook{X},\ehook{X}\rangle_\m$).
Informally, $\e$ can be regarded as a nondegenerate cross section of
the Klein bundle $(T^* M\otimes\m,H)$ over $\map(x)$,
where $\m \simeq \Rnum^n$.
This bundle is a reduction \cite{KobayashiNomizu}
of the orthonormal frame bundle of $M$, 
with the structure group $SO(n)$ reduced to $H$.
(In fact, $H \subseteq SO(n)$ with equality holding only when 
$M=G/H$ is a constant curvature manifold, 
$M =S^n \simeq SO(n+1)/SO(n)$ in the compact case.)

Recall \cite{Helgason},
the vector space $\m=\g/\h$ has the algebraic structure of
a union of orbits generated by the group $H^*$ acting on 
any fixed Cartan space (a maximal abelian subspace)
$\a \subset \m$ (all of which are isomorphic). 
Every such subspace $\a$, moreover, is a finite union of 
closed convex sets $\a_*(\m) \subseteq \a$ that are mutually isomorphic
under the Weyl group $W \subset H^*$ leaving $\a$ invariant.
(Up to closure, these sets $\a_*$ in $\a$
each can be identified with Weyl chambers \cite{Helgason}
associated to the set of positive roots of the Lie algebra $\g$
relative to $\m$.)

Since $H^*$ acts as a group of local gauge transformations 
on $\e$ and $\conx$,
a gauge equivalent coframe exists such that,
at every point $x$ along $\map$,
$\ehook{X}$ is a constant unit vector that 
lies in the (fixed) Cartan space $\a \subset \m$
on which the Weyl group acts as a discrete gauge equivalence group. 
This leads to the following algebraic characterization for curves $\map(x)$.

{\bf Proposition~3.2: }{\it
The constant unit vector $\ehook{X} \subset \a$ 
defines an algebraic invariant of the curve $\map$. 
Thus, all curves on $M=G/H$ can be divided into algebraic classes
in one-to-one correspondence with the unit-norm elements 
in the closed convex set $\a_*(\m) \simeq \a/W$ 
as determined by $\ehook{X}$. 
}

For a given vector $\ehook{X}$, 
a linear coframe $\e$ along $\map$ is determined only up to 
local gauge transformations \eqref{gaugetransf}
given by the group $\isoH \subset H^*$
that preserves $\ehook{X} \subset \a$.
Such transformations will be referred to as the 
{\it isotropy gauge group} of $\map$.
The isotropy gauge freedom can be used to adapt $\e$ and $\conx$
to $\map$ in a preferred way,
generalizing the algebraic notion of a parallel moving frame 
for curves in Riemannian geometry stated in Proposition~3.1.

{\bf Definition~3.3: }
A $\m$-valued linear coframe $\e$ along a curve $\map(x)$ 
in a Riemannian symmetric space $M=G/H$ is said to be 
{\it $H$-parallel} if its associated $\h$-valued linear connection 
$\conxhook{X}$ in the tangent direction $X=\mapder{x}$ of the curve
lies in the perp space $\hcenperp$ of the Lie subalgebra
$\hcen \subset\h$ of the isotropy gauge group 
$\isoH \subset H^*$ that preserves $\ehook{X}$
given by a fixed unit vector in $\a\subset \m$. 

To explain the geometrical meaning of such a coframe,
first decompose the dual frame $\duale=\duale_\parallel+\duale_\perp$
and the connection $\conx=\conx^\parallel+\conx^\perp$
according to 
\EQ\label{mhdecomp}
\m = \mcen \oplus \mcenperp ,\quad 
\h = \hcen \oplus \hcenperp ,
\endEQ
relative to the unit vector $\ehook{X} \subset \a\subset \m$.
The $\m$-valued linear frame $\duale$ will provide 
a framing of the curve $\map$ in $M$
(namely a linear map from the vector space 
$\m^* \simeq \Rnum^n$ onto $T_x^* M$ over $\map(x)$)
once a basis expansion for $\m,\h$ is introduced. 
Note the condition for $\duale$ to be $H$-parallel is 
\EQ\label{Hparallel}
\conxhook{X}^\parallel := \conx^\parallel\hook{X} =0
\endEQ
along with
\EQ\label{Hadapted}
\Dx \ehook{X} =0 ,\quad
|\ehook{X}|_\m =1 .
\endEQ

{\it Remark: }
This property $\conxhook{X}^\parallel=0$ is maintained by 
rigid ($x$-independent) gauge transformations 
in $\isoH$ that preserve $\ehook{X}$,
since under these transformations, 
$\Ad(h\inv)X=X$ implies $\Ad(h\inv)\conxhook{X}^\perp$
belongs to $\hcen$ 
and hence the transformed linear connection \eqref{gaugetransf}
satisfies $\tilde\conx_X^\parallel=0$. 
All such transformations comprise the {\it equivalence group},
denoted $\equivH$, 
for $H$-parallel moving frames. 
The infinitesimal equivalence group is simply given by 
the centralizer Lie subalgebra $\hcen \subset\h$,
as seen from the Lie bracket relations 
\eqrefs{perpspbracketm}{perpspbracketh}. 
Note $\equivH$ preserves the decomposition \eqref{mhdecomp}
of the tangent space $T_\map M\simeq_G =\m$
into algebraically parallel and perpendicular parts 
with respect to $X$. 

Let $\{\mvec{a}_\parallel,\hvec{i}_\parallel\}$
be respective orthonormal bases for the subspaces 
$\mcen$, $\hcen$, so 
\EQ\label{mhparbasis}
\ad(\ehook{X})\mvec{a}_\parallel 
= \ad(\ehook{X})\hvec{i}_\parallel
=0
\endEQ
(where $a=1,\ldots,\dim\mcen$; $i=1,\ldots,\dim\hcen$).
Recall, the perp subspaces $\mcenperp$, $\hcenperp$ 
are isomorphic under the linear map $\ad(\ehook{X})$,
and so let $\{\mvec{a}_\perp,\hvec{i}_\perp\}$
be respective orthonormal bases for 
$\mcenperp \simeq \hcenperp$
related by 
$\mvec{a}_\perp=\ad(\ehook{X})\inv\hvec{i}_\perp $
(where $a,i=1,\ldots,\dim\mcenperp=\dim\hcenperp$).
Orthonormality of each basis is given by 
\EQs
&&
\langle \mvec{a}_\parallel,\mvec{b}_\parallel\rangle_\m 
=\langle \mvec{a}_\perp,\mvec{b}_\perp\rangle_\m 
= -\delta^{ab} ,
\\
&&
\langle \hvec{i}_\parallel,\hvec{j}_\parallel\rangle_\h  
=\langle \hvec{i}_\perp,\hvec{j}_\perp\rangle_\h  
= -\delta^{ij} , 
\endEQs
and
\EQ\label{mhperpparbasis}
\langle \mvec{a}_\perp,\mvec{b}_\parallel\rangle_\m 
=\langle \hvec{i}_\perp,\hvec{j}_\parallel\rangle_\h  
=0 . 
\endEQ

Now, write 
\EQ
(\duale_\parallel)_a = -\langle \mvec{a}_\parallel,\duale\rangle_\m ,\qquad
(\duale_\perp)_a = -\langle \mvec{a}_\perp,\duale\rangle_\m , 
\endEQ
and 
\EQ
(\conx^\parallel)_i = -\langle \hvec{i}_\parallel,\conx\rangle_\h ,\qquad
(\conx^\perp)_i = -\langle \hvec{i}_\perp,\conx\rangle_\h ,
\endEQ
yielding the basis expansion of $\duale,\conx$. 
As a consequence of properties \eqsref{mhparbasis}{mhperpparbasis},
the vectors $(\duale_\parallel)_a$ and $(\duale_\perp)_a$
will be said, respectively, to be {\it algebraically} 
{\it parallel} ($\parallel$) and {\it perpendicular} ($\perp$)
to the tangent vector $X=\mapder{x}$ along the curve. 
This framing is adapted to $\map$ if it contains $X$ 
as one of the vectors parallel to $X$, (say for $a=1$)
\EQ
(\duale_\parallel)_1 = X ,
\endEQ 
which can be achieved by letting $\mvec{1}_\parallel =\ehook{X}$
be one of the basis vectors of $\mcen$. 

{\bf Lemma~3.4: }{\it
The frame vectors $\{(\duale_\parallel)_a,(\duale_\perp)_a\}$
provide an orthonormal basis of $T_\map M$ over the curve $\map(x)$. 
Through the soldering formula \eqref{soldere},
their covariant derivatives in the tangent direction $X=\mapder{x}$
are determined by the connection matrices
$(\conxhook{X}^\perp)_i \c{ib}{a}$
with 
$(\conxhook{X}^\parallel)_i \c{ib}{a}=0$. 
}

The Lie bracket relation \eqref{perpspbracketm}
between $\hcenperp$ and $\mcen$
now leads to an explicit geometric interpretation of 
the $H$-parallel property. 
Note $Y$ is algebraically $\parallel$ to $X$
iff $\ad(\ehook{X})\ehook{Y}=[\ehook{X},\ehook{Y}]=0$
namely $\ehook{Y}:= \e\hook Y$ belongs to $\mcen$. 
Likewise $Y$ is algebraically $\perp$ to $X$
iff $\ehook{Y}$ belongs to $\mcenperp$. 

{\bf Theorem~3.5: }{\it
For any $H$-parallel frame along a curve $\map(x)$,
the covariant derivatives 
$\covder{x} (\duale_\parallel)_a$, 
$a=1,\ldots,\dim\mcen$, 
are algebraically $\perp$ to the tangent vector $X=\mapder{x}$.
The other covariant derivatives 
$\covder{x} (\duale_\perp)_a$, 
$a=1,\ldots,\dim\mcenperp$, 
are algebraically $\parallel$ to $X$
iff the Lie bracket relation 
$[\mcenperp,\hcenperp]_\g \subseteq \mcen$ 
holds,
which is always the case whenever the linear map $\ad(\ehook{X})^2$
is a multiple of the identity on the vector spaces 
$\mcenperp\simeq\hcenperp$
(cf Proposition~2.4). 
}

These geometric properties of an $H$-parallel frame 
are a strict generalization of the Riemannian case \eqref{parallelprop}
whenever $\dim\mcenperp >1$, 
or $[\mcenperp,\hcenperp]_\g \cap \mcenperp \neq 0$. 

There is a further geometric meaning for an $H$-parallel connection itself.
Recall, the principal normal along any affine-parametrized curve 
is the normal vector \cite{Guggenheimer}
\EQ
N=\covder{x} \mapder{x} . 
\endEQ

{\bf Proposition~3.6: }{\it 
In an $H$-parallel adapted frame along a curve $\map(x)$,
the principal normal is given by 
$N=\langle \ad(\ehook{X}) \conxhook{X}^\perp, \duale\rangle_\m$
with the frame components
$N^a = \e_a\hook N 
= (\conxhook{X}^\perp)_i \c{i1}{a} := (\conxhook{X}^\perp)_a$
where $\c{i1}{a}=\ad(\ehook{X}){}_a{}^i$ is an invertible matrix
($a,i=1,\ldots,\dim\mcenperp=\dim\hcenperp$).
Moreover, these components are differential covariants of the curve
with respect to the rigid equivalence group 
(of $x$-independent linear transformations) $\equivH \subset H^*$
(preserving the framing). 
Namely the set of components $(\conxhook{X}^\perp)_i \c{i1}{a}$
is invariantly defined by $\map(x)$ 
up to the covariant action of the group $\equivH$. 
}

An important final remark now is that 
the $H$-parallel condition \eqref{Hparallel}
on the linear connection $\conx$
can always be achieved by means of 
a suitable gauge transformation \eqref{gaugetransf}
as follows. 
Under the isotropy gauge group $\isoH$,
$\conxhook{X}^\parallel$ is transformed to 
$\tilde\conx_{X}^\parallel
= \Ad(h^{-1}) \conxhook{X}^\parallel +h^{-1}\Dx h$
where $h$ is an arbitrary smooth function 
from the curve $\map(x) \subset M$
into the Lie subgroup of $\isoH\subset H$, 
satisfying $\Ad(h^{-1})\ehook{X} =0$.
Condition \eqref{Hparallel} applied to $\tilde\conx_{X}^\parallel$
yields an ODE on $h(x)$,
$h^{-1}\Dx h = -\Ad(h^{-1})\conxhook{X}^\parallel$,
for which local existence of a solution is a standard result. 
(In particular, in the adjoint representation of $H$ and $\h$, 
this ODE becomes a linear homogeneous matrix differential equation
$\Dx\Ad(h) = -\Ad(h)\ad(\conxhook{X}^\parallel)$.)
This result is a straightforward generalization of the standard existence
of a parallel moving frame in the Riemannian setting. 

Proposition~3.7:
By means of a gauge transformation in the isotropy gauge group $\isoH$,
an $H$-parallel frame $\{(\duale_\parallel)_a,(\duale_\perp)_a\}$
exists for any curve $\map(x)$ in a given Riemannian symmetric space $G/H$
and is unique up to the equivalence group 
$\equivH \subset H^*$ 
consisting of rigid ($x$-independent) gauge transformations
that preserve the algebraic structure \eqsref{mhdecomp}{Hadapted}
of the framing along $\map(x)$.

\section{Derivation of bi-Hamiltonian operators from 
non-stretching curve flows}

Consider an arbitrary smooth flow $\map(t,x)$ of any smooth curve
in an $n$-dimensional Riemannian symmetric space $M=G/H$. 
Write $X=\mapder{x}$ for the tangent vector
and $Y=\mapder{t}$ for the evolution vector
along the curve. 
Provided the flow is transverse to the curve, 
namely $X$ and $Y$ are not parallel vectors, 
then $\map(t,x)$ will describe a smooth two-dimensional surface
immersed in the manifold $M$. 

The torsion and curvature equations of the Riemannian connection $\covder{}$
on $M$ restricted to this surface $\map(t,x)$ are given by 
\EQs
&&
\tors(X,Y) = \covder{x}\mapder{t} - \covder{t}\mapder{x} =0 , 
\\
&&
\riem(X,Y) =[\covder{x},\covder{t}] = -\ad_\m([\mapder{x},\mapder{t}]_\m) , 
\endEQs
where $[\cdot,\cdot]_\m$ is the Lie bracket on $\m=\g/\h$
with the canonical identification $\m \simeq_G T_x M$. 
There is a straightforward frame formulation for these equations
obtained from the pullback of 
the Cartan structure equations \eqref{cartaneqs}
for any $\m$-valued linear coframe $\e$ 
and $\h$-valued linear connection $\conx$ on the surface $\map(t,x)$:
\EQs
&&
\Dx\e_t - \Dt\e_x +[\conx_x,\e_t]_\g - [\conx_t,\e_x]_\g =0 , 
\label{ecartan}\\
&&
\Dx\conx_t - \Dt\conx_x +[\conx_x,\conx_t]_\h =-[\e_x,\e_t]_\m , 
\label{conxcartan}
\endEQs
where
\EQs
&&
\e_x:= \e\hook X =\e\hook\mapder{x} ,\quad
\e_t:= \e\hook Y =\e\hook\mapder{t} , 
\label{eflow}\\
&&
\conx_x:= \conx\hook X =\conx\hook\mapder{x} ,\quad
\conx_t:= \conx\hook Y =\conx\hook\mapder{t} .
\label{conxflow}
\endEQs
These structure equations coincide with 
the zero-curvature equation \eqref{zerocurveq}
for $G$ viewed as a $H$-bundle
in a local trivialization over the surface $\map(t,x)$ in $M$. 
In particular, in local coordinates for $G \approx U\times H$,
the $\g$-valued connection $1$-form $\e+\conx =\gconx$
coincides with the pullback of the left invariant Maurer-Cartan form on $G$
to $U|_{\map} \subset M$,
locally giving a soldering of the frame geometry of $G/H$ onto
the Riemannian geometry of the manifold $M$.

\subsection{Non-stretching geometric curve flows}

A curve flow $\map(t,x)$ is {\it non-stretching} (inextensible) 
if it preserves the arclength $ds =|\mapder{x}|dx$ at every point
on the curve, namely $\Dt|\mapder{x}|=0$ 
or equivalently $\covder{t}\mapder{x}$ $(=\covder{x}\mapder{t})$ 
is orthogonal to $\mapder{x}$ in the Riemannian metric $g$ on $M$. 
Now let $\map(t,x)$ be a non-stretching curve flow 
given in terms of an affine parametrization $|\mapder{x}|=1$,
without loss of generality, where $x$ measures arclength along the curve. 
For any such flow, there exists an $H$-parallel frame formulation
obtained through Proposition~3.7. 
In particular, a suitable local gauge transformation in $H^*\simeq \Ad(H)$
(given by a change of coordinates for the local trivialization of $G$)
can be applied to the linear coframe $\e$ and linear connection $\conx$
to produce an $H$-parallel moving frame along each curve in the flow
$\map(t,x)$, depending smoothly on $t$.

{\bf Lemma~4.1: }{\it
In an $H$-parallel frame formulation for non-stretching curve flows, 
the projections \eqrefs{eflow}{conxflow} of $\e$ and $\conx$ 
in the tangent direction $X=\mapder{x}$ and the flow direction $Y=\mapder{t}$
have the following properties:
\vskip0pt
(i) $\e_x$ belongs to some Cartan space $\a \subset\m$, 
is constant ($\Dx\e_x=0$), and has the normalization $|\e_x|_\m=1$;
\vskip0pt
(ii) $\e_t:= h_\parallel +h_\perp$ 
and $\omega_t:=\varpi^\parallel+ \varpi^\perp$
each decompose into algebraically parallel and perpendicular parts
relative to $\e_x$,
namely $h_\parallel$ and $\varpi^\parallel$ 
belong to the centralizer subspaces
$\m_\parallel :=\mcen$ and $\h_\parallel :=\hcen$
while $h_\perp$ and $\varpi^\perp$ 
belong to the perp spaces
$\m_\perp:=\mcenperp$ and $\h_\perp:=\hcenperp$;
\vskip0pt
(iii) $\u:=\omega_x$ belongs to the perp space $\h_\perp$;
\vskip0pt
(iv) $h_\perp$, $\e_x$, $\u$ respectively determine 
the perpendicular flow vector 
$Y_\perp=(\mapder{t})_\perp=-\langle h_\perp,\duale\rangle_\m$, 
the tangent vector 
$X=\mapder{x}=-\langle \e_x,\duale\rangle_\m$, 
and the principal normal vector 
$N=\covder{x}\mapder{x}=-\langle \ad(\ehook{x})\u,\duale\rangle_\m$
for the curve. 
}

The Lie bracket relations in Proposition~2.5 
on the spaces 
$\m=\m_\parallel\oplus\m_\perp$
and $\h=\h_\parallel\oplus\h_\perp$
lead to a corresponding decomposition of 
the Cartan structure equations \eqrefs{ecartan}{conxcartan}
expressed in terms of the variables 
$\e_x,h_\parallel,h_\perp,\varpi^\parallel,\varpi^\perp,\u$. 
The torsion equation becomes
\EQs
&&
0= \Dx h_\parallel +[\u,h_\perp]_\parallel , 
\\
&&
0= \ad(\e_x)\varpi^\perp +\Dx h_\perp +[\u,h_\parallel] +[\u,h_\perp]_\perp , 
\endEQs
while the curvature equation becomes
\EQs
&&
0= \Dx \varpi^\parallel +[\u,\varpi^\perp]_\parallel , 
\\
&&
\ad(\e_x) h_\perp = 
\Dt\u -\Dx \varpi^\perp +[\varpi^\parallel,\u] +[\varpi^\perp,\u]_\perp , 
\endEQs
where $[\cdot,\cdot]_\parallel$ denotes the restriction of
the Lie bracket on $\g$ to $\m_\parallel$ or $\h_\parallel$,
and likewise $[\cdot,\cdot]_\perp$ denotes the restriction to
$\m_\perp$ or $\h_\perp$. 
Note that, in the present notation, 
the Lie bracket relations consist of 
\EQs
&&
[\m_\parallel,\m_\parallel] \subseteq \h_\parallel ,\quad
[\h_\parallel,\m_\parallel] \subseteq \m_\parallel ,\quad
[\h_\parallel,\h_\parallel] \subseteq \h_\parallel , 
\\
&&
[\m_\parallel,\m_\perp] \subseteq \h_\perp ,\quad
[\h_\parallel,\m_\perp] \subseteq \m_\perp ,\quad
[\h_\perp,\m_\parallel] \subseteq \m_\perp ,\quad
[\h_\parallel,\h_\perp] \subseteq \h_\perp , 
\endEQs
so thus 
\EQ\label{commperp}
[\m_\parallel,\m_\parallel]_\perp= [\h_\parallel,\m_\parallel]_\perp
= [\h_\parallel,\h_\parallel]_\perp =0 , 
\endEQ
and 
\EQ\label{commpar}
[\m_\parallel,\m_\perp]_\parallel = [\h_\parallel,\m_\perp]_\parallel
= [\h_\perp,\m_\parallel]_\parallel = [\h_\parallel,\h_\perp]_\parallel =0 . 
\endEQ
Therefore the only brackets with a possibly nontrivial decomposition are 
$[\m_\perp,\m_\perp]$ and $[\h_\perp,\h_\perp]$. 

{\bf Lemma~4.2: }{\it
The Cartan structure equations for any $H$-parallel 
linear coframe $\e$ and linear connection $\conx$ 
pulled back to the two-dimensional surface $\map(t,x)$ 
determine a flow on $\u=\conx_x$ given by 
\EQs
&&
\u_t
= \Dx \varpi^\perp +[\u,\varpi^\parallel] + [\u,\varpi^\perp]_\perp
+\ad(\e_x) h_\perp , 
\label{uteq}\\
&&
\varpi^\perp 
= -\ad(\e_x)\inv( \Dx h_\perp +[\u,h_\parallel] +[\u,h_\perp]_\perp ) , 
\label{wperpeq}
\endEQs
with 
\EQ\label{hparwpareq}
h_\parallel = -\Dinvx [\u,h_\perp]_\parallel ,\quad
\varpi^\parallel = -\Dinvx [\u,\varpi^\perp]_\parallel , 
\endEQ
where $\Dinvx$ denotes the formal inverse of 
the total $x$-derivative operator $\Dx$. 
}

Note that, once $h_\perp$ is specified, 
this determines both
the flow on $\u$ and the curve flow of $\map$ itself,
where $h_\parallel,\varpi^\parallel,\varpi^\perp$ are given 
in terms of $h_\perp$ by equations \eqsref{wperpeq}{hparwpareq}.
A natural class of geometric flows will now be introduced.

{\bf Definition~4.3: }
A curve flow is said to be {\it geometric} if
$h_\perp$ is an equivariant $\m_\perp$-valued function of
$(x,\u,\u_x,\ldots)$ under the action of the rigid ($x$-independent) group of
linear transformations $\equivH$,
where the arclength variable $x$ is an invariant of the curve,
and the components of flow variable $\u$ are differential covariants of
the curve (cf Proposition~3.6).

In the case of polynomial geometric flows, $\m_\perp$-equivariance means that
the function $h_\perp$ is constructible in terms of 
$(\u,\u_x,\ldots)$ together with $\e_x$ and $x$
using just the Lie bracket and Cartan-Killing inner product 
on these variables.
Such flows correspond to curve flows $\map(t,x)$ such that
$\mapder{t}=f(x,\mapder{x},\mapder{xx},\ldots)$ is constructed using
the metric $g(\cdot,\cdot)$ and the linear map $\ad_x(\cdot)^2$
on $T_x M \simeq_G \m$ via the relations (i) to (iv) in Lemma~4.1. 

{\bf Proposition~4.4: }{\it
For a given polynomial geometric flow on $\u$,
the curve $\map$ obeys a $G$-invariant flow equation in $M$
determined through the identifications
\EQ\label{idents}
\mapder{x} \leftrightarrow \e_x ,\quad
\covder{x} \leftrightarrow \Dx+[\u,\cdot]_\m :=\frder_x ,\quad
\ad_x(\cdot) \leftrightarrow \ad_\m(\cdot) ,\quad
g(\cdot,\cdot)\leftrightarrow -\langle\cdot,\cdot\rangle_\m .
\endEQ
}

\subsection{Hamiltonian operators and bi-Hamiltonian structure}

As a main result, the flow equation on $\u$ 
will now be shown to possess an elegant 
bi-Hamiltonian structure. 
To begin, recall that $\ad(\ehook{x})$ as a linear map 
gives an isomorphism between the spaces $\m_\perp$ and $\h_\perp$.
Write $h^\perp:=\ad(\ehook{x})h_\perp$ for the image of $h_\perp$
which belongs to the same perp space $\h_\perp$ as $\u$.
Expressing $\u$ in terms of $h^\perp$ and $\varpi^\perp$ 
then leads to the following operator form for the flow equation \eqref{uteq}:
\EQ\label{ufloweq}
\u_t = \Hop(\varpi^\perp) +h^\perp ,\quad
\varpi^\perp = \Jop(h^\perp) , 
\endEQ
where 
\EQ
\Hop = {\mathcal K}|_{\h_\perp} ,\quad
\Jop= -\ad(\ehook{x})\inv {\mathcal K}|_{\m_\perp} \ad(\ehook{x})\inv
\label{HJops}
\endEQ
are linear operators which act on $\h_\perp$-valued functions
and are invariant under $\equivH$,
as defined in terms of the operator
\EQ\label{Kop}
\Kop := 
\Dx +[\u,\cdot]_\perp -[\u,\Dinvx[\u,\cdot]_\parallel] . 
\endEQ

{\bf Theorem~4.5: }{\it
$\Hop$, $\Jop$
are compatible Hamiltonian cosymplectic and symplectic operators
with respect to the $\h_\perp$-valued flow variable $\u$.
For geometric flows given by $h^\perp(x,\u,\u_x,\ldots)$, 
the flow equation on $u$ has 
an $\equivH$-invariant form $\u_t = \Hop_1(\Jop(h^\perp))$
where $\Hop_1=\Hop +\Jop\inv$ is a formal Hamiltonian operator. 
Consequently, such a geometric flow will have a Hamiltonian structure
\EQ
\u_t = \Hop_1(\delta\ham{H}/\delta\u)
\endEQ
for some Hamiltonian functional $\ham{H}=\int H(x,\u,\u_x,\ldots) dx$ iff
\EQ
\delta\ham{H}/\delta\u = \varpi^\perp = \Jop(h^\perp) . 
\endEQ
Moreover, all such Hamiltonian flows have a second Hamiltonian structure 
\EQ
\u_t = \Hop_2(\delta\ham{F}/\delta\u)
\endEQ
given by the (compatible) 
formal Hamiltonian operator $\Hop_2=\Rop\inv\Hop_1$,
with a Hamiltonian functional $\ham{F}=\int F(x,\u,\u_x,\ldots) dx$
such that
\EQ
\delta\ham{F}/\delta\u = \Rop^*(\varpi^\perp) = \Jop(\Rop(h^\perp)) , 
\endEQ
where $\Rop=\Hop\Jop$ is a (hereditary) recursion operator
(and $\Rop^*=\Jop\Hop$ is its adjoint).
}

Here $\delta\ham{H}/\delta\u$ denotes the variational derivative of
the real-valued functional $\ham{H}$. 
More precisely, $\delta\ham{H}/\delta\u$ is defined as the unique 
$\h_\perp$-valued function of $(x,\u,\u_x,\ldots)$ such that
\EQ
-\int\langle \delta\ham{H}/\delta\u,h^\perp\rangle dx 
= \delta_{h^\perp}\ham{H}
\endEQ
formally holds for all $\h_\perp$-valued functions $h^\perp$, 
where $\delta_{h^\perp} = \pr(h^\perp\cdot\partial/\partial\u)$
denotes the variation induced by $h^\perp$, namely the prolongation of
the given vector field $h^\perp\cdot\partial/\partial\u$ on $J^\infty$,
and where $\langle\cdot,\cdot\rangle$ 
denotes the (negative-definite) Cartan-Killing inner product on $\h_\perp$. 
(A dot will denote summation over Lie-algebra components
with respect to any fixed basis.)

The definition of cosymplectic and symplectic operators 
and a proof of this theorem will be given in sections~4.3--4.5.
A summary of the basic theory of bi-Hamiltonian operators
and Hamiltonian structures for scalar evolution equations
is given in \cite{Olver,Dorfman}. 
The presentation here will extend this theory to a Lie-algebra valued setting. 

One main property characterizing Hamiltonian operators consists of 
the symplectic structure 
they induce on the $x$-jet space $(x,\u,\u_x,\ldots):= J^\infty$ 
of the flow variable $\u$.
In particular, 
there is associated to $\Hop$ the Poisson bracket defined by 
\EQ\label{PB}
\{\ham{H},\ham{G}\}_\Hop 
= -\int\langle \Hop(\delta\ham{G}/\delta\u),\delta\ham{H}/\delta\u\rangle  dx
\endEQ
for any pair of functionals $\ham{H},\ham{G}$.
This bracket is skew-symmetric
\EQ\label{PBskew}
\{\ham{H},\ham{G}\}_\Hop= -\{\ham{G},\ham{H}\}_\Hop
\endEQ
and obeys the Jacobi identity
\EQ\label{PBjacobi}
\{\ham{H},\{\ham{G},\ham{F}\}_\Hop\}_\Hop
+\{\ham{G},\{\ham{F},\ham{H}\}_\Hop\}_\Hop
+\{\ham{F},\{\ham{H},\ham{G}\}_\Hop\}_\Hop
=0
\endEQ
as a consequence of $\Hop$ being cosymplectic. 
Let $h^\perp(x,\u,\u_x,\ldots)$ be a function with values 
in the space $\h_\perp$.
Then, on the $x$-jet space $J^\infty$, 
$h^\perp\cdot\partial/\partial\u$ is a {\it Hamiltonian vector field}
with respect to $\Hop$ if there exists a (Hamiltonian) functional $\ham{H}$
such that 
\EQ
\delta_{h^\perp}\ham{G} = \{\ham{G},\ham{H}\}_\Hop
\endEQ
for all functionals $\ham{G}$. 
Associated to $\Jop$ is a symplectic $2$-form defined by 
\EQ\label{sympl2form}
\sympform
(h_1^\perp\cdot\partial/\partial\u,h_2^\perp\cdot\partial/\partial\u)_\Jop
= -\int\langle h_1^\perp,\Jop(h_2^\perp)\rangle  dx
\endEQ
for any pair of Hamiltonian vector fields given by $h_1^\perp,h_2^\perp$.
The property that $\sympform$ is symplectic
corresponds to the skew-symmetry and Jacobi identity of
the Poisson bracket $\{\cdot,\cdot\}_{\Jop\inv}$
defined using the formal inverse of $\Jop$. 
(Consequently, 
$\Jop\inv$ is formally cosymplectic,
and conversely $\Hop\inv$ is formally symplectic.)
This $2$-form $\sympform$
gives a canonical pairing between Hamiltonian vector fields
and corresponding covector fields defined as follows. 
Let $\varpi^\perp(x,\u,\u_x,\ldots)$ be a function with values 
in the space $\h_\perp$. 
Then the covector field $\varpi^\perp\cdot d\u$ is dual to 
a Hamiltonian vector field $h^\perp\cdot\partial/\partial\u$ if
\EQ
\varpi^\perp\cdot d\u := 
\sympform(\cdot\ ,h^\perp\cdot\partial/\partial\u)_\Jop
\endEQ
holds with respect to $\Jop$.
As a consequence of $\Jop$ being symplectic,
such covector fields are {\it variational}, namely there exists
a (Hamiltonian) functional $\ham{E}$ such that
\EQ
\delta_{h^\perp}\ham{E} 
= -\int\langle \varpi^\perp,h^\perp\rangle dx
\endEQ
for all vector fields $h^\perp\cdot\partial/\partial\u$. 

{\bf Proposition~4.6: }{\it
The Hamiltonian (cosymplectic and symplectic) operators \eqref{HJops}
give mappings 
$h^\perp\cdot\partial/\partial\u 
\rightarrow 
\Jop(h^\perp)\cdot d\u$
and
$\varpi^\perp\cdot d\u 
\rightarrow 
\Hop(\varpi^\perp)\cdot \partial/\partial\u$
between Hamiltonian vector fields and variational covector fields
on the $x$-jet space of the flow variable $\u$. 
}

Theorem~4.5 and Proposition~4.6 provide a broad generalization of
known results on the geometric origin of bi-Hamiltonian structures
in Riemannian symmetric spaces $M=G/H$. 
Special cases were first derived in \cite{SandersWang1,Anco1}
for constant-curvature spaces $SO(N+1)/SO(N) \simeq S^N$
and subsequently generalized in \cite{sigmapaper} to the spaces $G/SO(N)$,
covering all examples in which the bi-Hamiltonian structure
is a $O(N-1)$-invariant 
(namely, $\equivH = O(N-1)\subset SO(N)$). 
The proof of Theorem~4.5 in the general case is similar to 
the special case proven in \cite{SandersWang1}
where $\h_\parallel =\alg{so}(N-1)\simeq \Rnum^{N-1}\wedge\Rnum^{N-1}$, 
$\h_\perp\simeq \Rnum^{N-1}$
such that 
$[\h_\perp,\h_\parallel]\subseteq \h_\perp$
and
$[\h_\perp,\h_\perp]=\h_\parallel$. 
Note these Lie bracket relations correspond to 
the geometrical properties \eqref{riemparallel} 
of a Riemannian parallel frame \cite{sigmapaper}. 
There are two main differences in the general case. 
Firstly, 
the operators $\Hop,\Jop$ have an additional term arising from 
the Lie bracket structure 
$[\h_\perp,\h_\perp]\cap \h_\perp \neq 0$
as permitted by the more general algebraic notion of a parallel frame
in Definition~3.3 and Theorem~3.5.
Secondly, 
these operators are formulated here in a manifestly 
$\equivH$-invariant fashion, 
using just the intrinsic Lie bracket on $\h=\h_\parallel\oplus\h_\perp$,
as opposed to a more cumbersome $\Rnum^{N-1}$-component form 
used in \cite{SandersWang1}.

\subsection{Proof of cosymplectic property for $\Hop$}

A linear operator $\Hop$ that acts on $\h_\perp$-valued functions
in the $x$-jet space $J^\infty$ of the flow variable $\u$
is a {\it Hamiltonian cosymplectic operator} \cite{Olver,Dorfman} iff
its associated Poison bracket \eqref{PB}
is skew-symmetric and obeys the Jacobi identity. 

Skew-symmetry \eqref{PBskew} is equivalent to $\Hop$ being skew-adjoint
with respect to the natural inner product 
$\int\langle f_1,f_2\rangle dx$ 
on $\h_\perp$-valued functions $f$ on $J^\infty$ 
induced by the Cartan-Killing inner product on $\h_\perp$. 
This condition can be formulated in a pointwise manner on $J^\infty$
as follows:
$\Hop$ is skew-adjoint iff
\EQ\label{Hskew}
\langle f,\Hop(f)\rangle \equiv 0 \text{ modulo a total $x$-derivative }
\endEQ
(namely, $\langle f,\Hop(f)\rangle = \Dx\Upsilon$
for some scalar function $\Upsilon$). 

{\bf Proposition~4.7: }{\it
The linear operator 
$\Hop = \Dx +[\u,\cdot]_\perp -[\u,\Dinvx[\u,\cdot]_\parallel]$
in Theorem~4.5 is skew-adjoint. 
}

Proof. 
The l.h.s. of \eqref{Hskew} is given by 
\EQ
\langle f,\Dx f+[u,f]_\perp -[u,\Dinvx [u,f]_\parallel] \rangle
\equiv
\langle f,[u,f]\rangle -\langle f,[u,\Dinvx [u,f]_\parallel] \rangle
\endEQ
since $\langle f,\Dx f\rangle=-\frac{1}{2}\Dx|f|^2$.
The remaining two terms can be simplified by means of the identities
\EQ
\langle A,[B,C_\perp] \rangle = \text{cyclic} ,\quad
\langle A,[B,C_\parallel] \rangle = \langle C_\parallel ,[A,B]\rangle , 
\endEQ
for functions $A,B$ with values in $\h_\perp$
and $C$ in $\h$. 
Thus
\EQ
\langle f,[u,f]\rangle 
= -\langle u,[f,f]\rangle =0
\endEQ
and 
\EQ
\langle f,[u,A(f)] \rangle
= -\langle A(f),[u,f]_\parallel \rangle
= \frac{1}{2}\Dx|A(f)|^2
\equiv 0
\endEQ
where $A(f) = \Dinvx [u,f]_\parallel$. 
Hence the l.h.s. of \eqref{Hskew} reduces to a total $x$-derivative,
which completes the proof. 

For a skew-adjoint operator $\Hop$, the Jacobi property \eqref{PBjacobi}
is well-known to be equivalent to the vanishing of the Schouten bracket of
$\Hop$ with itself \cite{Dorfman}.
The simplest approach to verifying if an operator 
has vanishing Schouten bracket 
is the calculus of multi-vectors developed in \cite{Olver}
which will be adapted to the $\h_\perp$-valued setting here. 

First, on $x$-jet space $J^\infty$,
introduce a $\h_\perp$-valued {\it vertical uni-vector} $\uni$
which is dual to $d\u$ regarded as a $\h_\perp$-valued 1-form,
in the sense that 
$-\langle \uni,d\u \rangle=1$
where the pairing denotes the usual contraction between vectors and 1-forms,
combined with their Cartan-Killing inner product on $\h$. 
Then $x$-derivatives of $\uni$ are introduced similarly as duals to 
$d\u_x=\Dx d\u$, $d\u_{xx}=\Dx^2 d\u$, \etc/
via 
$\uni_x=\Dx\uni$, $\uni_{xx}=\Dx^2\uni$, \etc/
such that 
$-\langle \uni_{kx},d\u_{lx}\rangle=\delta_{kl}$
(with $k,l=0,1,2,\ldots$). 
Formal sums of antisymmetric tensor products of these vertical uni-vectors
will define {\it vertical multi-vectors},
for instance $\uni\wedge\uni$ and $\uni\wedge\uni\wedge\uni$ 
with respective values in $\Lambda^2_{\h_\perp}$ and $\Lambda^3_{\h_\perp}$. 
More particularly, 
a combined wedge product and Lie bracket of $\uni$ with itself
yields a $\h$-valued vertical bi-vector
\EQ\label{bivec}
[\uni,\uni] = [\uni,\uni]_\perp + [\uni,\uni]_\parallel
\endEQ
while the vertical tri-vector 
\EQ\label{trivec}
[[\uni,\uni],\uni] = 0
\endEQ
vanishes due to the Jacobi identity on $\h$. 

It is useful to let the basic uni-vector $\uni$ have 
the formal action of a vector field on $\h_\perp$-valued functions 
$f(x,\u,\u_x,\ldots)$ on $J^\infty$, 
\EQ
\univar f := \sum_{0\leq k} \uni_{kx} \partial f/\partial u_{kx} . 
\endEQ
Likewise, let $\uni$ act on differential operators 
$\op=\sum_{0\leq k} a_k \Dx^k$ 
by a formal Lie derivative
\EQ
\univar \op := \sum_{0\leq k} (\univar a_k) \Dx^k
\endEQ
in terms of coefficients $a_k(x,\u,\u_x,\ldots)$ 
given by functions on $J^\infty$. 
In particular, $\univar\Dx=0$. 
Finally, the action of $\uni$ can be naturally extended first to 
vertical uni-vectors $\uni$ via defining 
\EQ
\univar\uni :=0
\endEQ
so $\univar \uni_{kx}=0$,
and then to vertical multi-vectors by letting $\univar$ 
act as a derivation. 

In this formalism the Schouten bracket of $\Hop$ with itself
vanishes if and only if
$\int\langle\uni\wedge \uniopvar{\Hop}\Hop(\uni)\rangle dx=0$
where $\langle\cdot\wedge\cdot\rangle$ 
denotes the Cartan-Killing inner product combined with the wedge product
of $\h_\perp$-valued 1-forms on $J^\infty$. 
Note, in particular, $\langle\uni\wedge\uni\rangle=0$. 
An equivalent pointwise formulation is that 
$\langle\uni\wedge \uniopvar{\Hop}\Hop(\uni)\rangle = \Dx\Upsilon$
(for some scalar function $\Upsilon$) or simply
\EQ
\langle\uni\wedge \uniopvar{\Hop}\Hop(\uni)\rangle \equiv 0
\text{ modulo a total $x$-derivative. }
\label{cosympeq}
\endEQ
This condition is equivalent to the Jacobi identity \eqref{PBjacobi} 
holding for the Poisson bracket $\{\cdot,\cdot\}_\Hop$. 

{\bf Proposition~4.8: }{\it
The linear operator 
$\Hop = \Dx +[\u,\cdot]_\perp -[\u,\Dinvx[\u,\cdot]_\parallel]$
in Theorem~4.5
has vanishing Schouten bracket and hence is cosymplectic. 
}

The following multi-vector identities will be used in the proof:
\EQs
&&
[A(\uni),B(\uni)]=[B(\uni),A(\uni)] , 
\label{symmcomm}\\
&&
[A(\uni),[B(\uni),C(\uni)]] + \text{cyclic} 
=0 , 
\label{jacobicomm}
\endEQs
and
\EQs
&&
\langle A(\uni)\wedge B(\uni)\rangle 
= -\langle B(\uni)\wedge A(\uni)\rangle , 
\label{skewtr}\\
&&
\langle A(\uni)\wedge [B(\uni),C(\uni)]\rangle 
= \text{cyclic} , 
\label{cyclictr}
\endEQs
for linear operators $A,B,C$ from $\h_\perp$ into $\h$; 
also, 
\EQs
&&
\langle A(\uni)\wedge [\u,\cdot]\rangle 
= -\langle \u, [A(\uni),\cdot]\rangle 
= -\langle [\u,A(\uni)]\wedge \cdot\rangle , 
\label{utr}\\
&&
[[\u,B(\uni)],A(\uni)] + [[\u,A(\uni)],B(\uni)]
= [\u,[B(\uni),A(\uni)]] , 
\label{unijacobi}
\endEQs
and in particular
\EQ
[[\u,A(\uni)],A(\uni)]] 
= \frac{1}{2}[\u,[A(\uni),A(\uni)]] . 
\label{unibrac}
\endEQ

Proof. 
Consider
\EQ
\uniopvar{\Hop}\Hop(\uni) = 
[\Hop(\uni),\uni]_\perp -[\Hop(\uni),\Dinvx[\u,\uni]_\parallel]
-[\u,\Dinvx[\Hop(\uni),\uni]_\parallel]
\label{uniHvarH}
\endEQ
which expands out to a $\h_\perp$-valued cubic polynomial in $\u$. 
There is a single inhomogeneous term, $[\Dx\uni,\uni]_\perp$.
This yields in the l.h.s of \eqref{cosympeq}:
\EQ
\langle\uni\wedge[\Dx\uni,\uni]\rangle
= \frac{1}{3}\Dx\langle\uni\wedge[\uni,\uni]\rangle
\equiv 0
\endEQ
by use of \eqref{cyclictr}. 

Next the linear terms in \eqref{uniHvarH} are given by 
\EQ
-[\Dx\uni,\Dinvx[\u,\uni]_\parallel]
-[\u,\Dinvx[\Dx\uni,\uni]_\parallel]
+[[\u,\uni]_\perp,\uni]_\perp . 
\label{cosymplinterm}
\endEQ
In the l.h.s of \eqref{cosympeq} 
the first term of \eqref{cosymplinterm} yields:
\EQ
-\langle\uni\wedge[\Dx\uni,\Dinvx[\u,\uni]_\parallel]\rangle
= -\frac{1}{2}\langle\Dinvx[\u,\uni]_\parallel\wedge\Dx[\uni,\uni]\rangle
\equiv \frac{1}{2}\langle [\u,\uni]_\parallel\wedge [\uni,\uni]\rangle
\endEQ
from \eqrefs{cyclictr}{symmcomm}. 
Similarly
the middle term of \eqref{cosymplinterm} simplifies:
\EQs
&&
-\langle\uni\wedge [\u,\Dinvx[\Dx\uni,\uni]_\parallel]\rangle
= \langle \u,[\uni,\Dinvx[\Dx\uni,\uni]_\parallel]\rangle
= -\frac{1}{2} \langle \u,[\uni,[\uni,\uni]_\perp]\rangle
\nonumber\\
&&
= -\frac{1}{2} \langle [\u,\uni]\wedge [\uni,\uni]_\perp\rangle
\endEQs
from \eqref{utr}, \eqrefs{bivec}{trivec}. 
The last term of \eqref{cosymplinterm} directly gives:
\EQ
\langle \uni\wedge[[\u,\uni]_\perp,\uni] \rangle
= \langle [\u,\uni]_\perp \wedge [\uni,\uni]\rangle . 
\endEQ
Hence the linear terms in the l.h.s of \eqref{cosympeq} combine into
\EQ
\equiv
\frac{1}{2}\langle [\u,\uni]_\parallel\wedge [\uni,\uni]\rangle
+ \frac{1}{2} \langle [\u,\uni]_\perp\wedge [\uni,\uni]\rangle
=\frac{1}{2} \langle [\u,\uni]\wedge [\uni,\uni]\rangle
= \frac{1}{2} \langle \u,[\uni,[\uni,\uni]]\rangle
=0
\endEQ
by \eqrefs{utr}{trivec}.

The quadratic terms in \eqref{uniHvarH} are given by 
\EQ
-[[\u,\uni]_\perp,\Dinvx[\u,\uni]_\parallel]
-[[\u,\Dinvx[\u,\uni]_\parallel],\uni]_\perp
-[\u,\Dinvx[[\u,\uni]_\perp,\uni]_\parallel] . 
\label{cosympquadrterm}
\endEQ
Write $A(\uni)= \Dinvx[\u,\uni]_\parallel$ 
and combine the first and second terms in \eqref{cosympquadrterm}
by the Jacobi relation \eqref{jacobicomm} to get
$[[A(\uni),\uni],\u]_\perp$. 
This yields in the l.h.s of \eqref{cosympeq}:
\EQ
\langle \uni\wedge [[A(\uni),\uni],\u]\rangle
= \langle A(\uni)\wedge[[\u,\uni],\uni]\rangle
\endEQ
from \eqrefs{utr}{cyclictr}. 
For the third term in \eqref{cosympquadrterm}, 
we use \eqref{utr} to simplify the l.h.s of \eqref{cosympeq}:
\EQ
-\langle\uni\wedge[\u,\Dinvx[[\u,\uni]_\perp,\uni]_\parallel]\rangle
= \langle[\u,\uni]\wedge\Dinvx[[\u,\uni]_\perp,\uni]_\parallel\rangle
\equiv -\langle A(\uni)\wedge[[\u,\uni]_\perp,\uni]\rangle . 
\endEQ
This term now combines with the previous one, giving
\EQ
\equiv
\langle A(\uni)\wedge[[\u,\uni]_\parallel,\uni]\rangle
= \langle A(\uni)\wedge[\Dx A(\uni),\uni]\rangle
= \frac{1}{2}\langle \uni\wedge\Dx[A(\uni),A(\uni)]_\perp\rangle
\label{combquadrcosymp}
\endEQ
from \eqref{cyclictr}. 
But the Lie bracket relation \eqref{commperp}
implies $[A(\uni),A(\uni)]_\perp$ vanishes 
and hence so does the combined quadratic term \eqref{combquadrcosymp}
in the l.h.s of \eqref{cosympeq}. 

Finally, there are two cubic terms in \eqref{uniHvarH}:
\EQ
[[\u,A(\uni)],A(\uni)]
+[\u,\Dinvx[[\u,A(\uni)],\uni]_\parallel]
\label{cosympcubterm}
\endEQ
where $A(\uni)= \Dinvx[\u,\uni]_\parallel$ again. 
The second term of \eqref{cosympcubterm}
simplifies in the l.h.s of \eqref{cosympeq}:
\EQs
&&
\langle \uni\wedge[\u,\Dinvx[[\u,A(\uni)],\uni]_\parallel]\rangle
= -\langle [\u,\uni]\wedge\Dinvx[[\u,A(\uni)],\uni]_\parallel\rangle
\nonumber\\
&&
\equiv \langle A(\uni)\wedge [[\u,A(\uni)],\uni]\rangle
= \langle \uni\wedge [A(\uni),[\u,A(\uni)]]\rangle
\endEQs
by \eqrefs{utr}{cyclictr}. 
This term combines with the first term of \eqref{cosympcubterm}
in the l.h.s of \eqref{cosympeq}:
\EQs
&&
2\langle \uni\wedge [[\u,A(\uni)],A(\uni)]\rangle
= \langle \uni\wedge [\u,[A(\uni),A(\uni)]_\parallel]\rangle
= -\langle \Dx A(\uni)\wedge [A(\uni),A(\uni)]\rangle
\nonumber\\
&&
\equiv -\frac{1}{3}\Dx\langle A(\uni)\wedge [A(\uni),A(\uni)]\rangle
\equiv 0
\endEQs
by means of \eqref{unibrac}, \eqrefs{cyclictr}{utr}, 
in addition to $[A(\uni),A(\uni)]_\perp=0$. 

Thus the l.h.s of \eqref{cosympeq} vanishes
modulo total $x$-derivatives, 
which completes the proof.

\subsection{Proof of symplectic property for $\Jop$}

A linear operator $\Jop$ that acts on $\h_\perp$-valued functions
in the $x$-jet space $J^\infty$ of the flow variable $\u$
is a {\it symplectic operator} \cite{Dorfman} iff 
the associated bilinear form \eqref{sympl2form}
is skew and closed, 
namely 
$\sympform(X,X)=0$
and 
$d\sympform(X,Y,Z)=\delta\downindex{X} \sympform(Y,Z) + \text{cyclic} = 0$
for all vector fields $X,Y,Z$ on $J^\infty$. 

The conditions on $\sympform$ have a straightforward pointwise formulation
on $J^\infty$ in terms of 
\EQ
\omega_\Jop (h_1^\perp,h_2^\perp) 
:= -\langle h_1^\perp,\Jop(h_2^\perp)\rangle
\endEQ
where $h^\perp$ denotes a $\h_\perp$-valued function. 
It will be convenient to work equivalently 
with $\m_\perp$-valued functions
$h_\perp = \ad(\ehook{x})\inv h^\perp$
along with the operator 
$\Kop|_{\m_\perp} = -\ad(\ehook{x})\Jop \ad(\ehook{x})$
as follows:
\EQ
\omega_\Jop (h_1^\perp,h_2^\perp) 
= \langle \ad(\ehook{x})h_{\perp 1}, \ad(\ehook{x})\inv\Kop(h_{\perp2})\rangle
= -\langle h_{\perp 1},\Kop(h_{\perp 2})\rangle . 
\endEQ
Note, here, $\ad(\ehook{x})$ is skew-adjoint with respect to 
the Cartan-Killing inner product on $\h_\perp\oplus\m_\perp$
as a consequence of the elementary identity 
\EQ\label{adskew}
\langle \ad(\ehook{x})A, B\rangle 
= \langle \ehook{x},[A, B]\rangle 
= -\langle A,\ad(\ehook{x})B\rangle 
\endEQ
for functions $A,B$ with values in $\h_\perp\oplus\m_\perp$. 

To proceed, 
skew-symmetry of $\sympform$ will hold iff 
$\omega_\Jop$ is skew, 
which is clearly equivalent to $\Kop$ being a skew-adjoint operator:
\EQ
\langle f,\Kop(f)\rangle \equiv 0 \text{ modulo a total $x$-derivative }
\endEQ
for all $\m_\perp$-valued functions $f$ on $J^\infty$. 
This condition is formally identical to the skew-adjoint property 
established earlier for the operator $\Hop=\Kop|_{\h_\perp}$
in Proposition~4.7,
where, recall, $\Kop$ is the operator given in \eqref{Kop}. 
As a result, 
the isomorphism $\m_\perp \simeq \h_\perp$ provided by $\ad(\ehook{x})$
implies that the steps in the proof of Proposition~4.7
carry through with the function $f$ now being $\m_\perp$-valued 
instead of $\h_\perp$-valued. 

{\bf Proposition~4.9: }{\it
The linear operator 
$\Jop = -\ad(\ehook{x})\inv( 
\Dx +[\u,\cdot]_\perp -[\u,\Dinvx[\u,\cdot]_\parallel] )$
$\ad(\ehook{x})\inv$
in Theorem~4.5 is skew-adjoint
and hence the bilinear form 
$\sympform(\cdot,\cdot) = \int\omega_\Jop(\cdot,\cdot) dx$
is skew-symmetric. 
}

Next, for a skew-adjoint operator $\Jop$,
the closure property $d\sympform=0$ 
is equivalent to having $\omega_\Jop$ satisfy 
$\delta_{h_1^\perp} \omega_\Jop(h_2^\perp,h_3^\perp) + \text{cyclic}
= \Dx\Upsilon$
for some scalar function $\Upsilon$. 
This condition will hold iff:
\EQ
\langle h_{\perp 3}, \delta_{h_1^\perp} \Kop(h_{\perp 2}) \rangle 
+ \text{cyclic}
=0
\label{sympeq}
\endEQ
for all $\m_\perp$-valued functions $h_{\perp i}$,
with $h_i^\perp = \ad(\ehook{x})h_{\perp i}$
($i=1,2,3$). 

Hereafter it will be useful to let
\EQ
(\cdot,\cdot) := [\ad(\ehook{x})\cdot,\cdot]
\label{adbracket}
\endEQ
which satisfies the following identities:
\EQ
\langle A, (B,C) \rangle 
= -\langle C, (B,A) \rangle 
\label{adskewtr}
\endEQ
and also
\EQs
&&
(A,B)_\parallel = (B,A)_\parallel , 
\label{adsymm}\\
&&
(A,B)_\perp- (B,A)_\perp = \ad(\ehook{x})[A,B] , 
\label{adsymmperp}\\
&&
\langle A, (B,C)_\perp \rangle + \text{cyclic} =0 , 
\label{adcyclictr}
\endEQs
for functions $A,B,C$ with values in $\m_\perp$.
Note the latter two identities are readily established from the property that 
$\ad(\ehook{x})$ acts as a derivation on $[\cdot,\cdot]$ in $\m_\perp$
and annihilates $\m_\parallel$. 

{\bf Proposition~4.10: }{\it
The skew bilinear form 
$\sympform(\cdot,\cdot) = \int\omega_\Jop(\cdot,\cdot) dx$
is closed
and hence the 
linear operator 
$\Jop = -\ad(\ehook{x})\inv( 
\Dx +[\u,\cdot]_\perp -[\u,\Dinvx[\u,\cdot]_\parallel] )\ad(\ehook{x})\inv$
in Theorem~4.5 is symplectic. 
}

Proof. 
Consider, from \eqrefs{Kop}{adbracket},
\EQs
\delta_{h_1^\perp} \Kop(h_{\perp 2}) 
&=& 
[h_1^\perp,h_{\perp 2}]_\perp
-[h_1^\perp,\Dinvx[\u,h_{\perp 2}]_\parallel]
-[\u,\Dinvx[h_1^\perp,h_{\perp 2}]_\parallel]
\nonumber\\
&=&
(h_{\perp 1},h_{\perp 2})_\perp
-(h_{\perp 1},\Dinvx[\u,h_{\perp 2}]_\parallel)
-[\u,\Dinvx(h_{\perp 1},h_{\perp 2})_\parallel]
\nonumber\\
&&
\label{varK}
\endEQs
which is a $\m_\perp$-valued linear polynomial in $\u$. 
The inhomogeneous term yields in the l.h.s of \eqref{sympeq}:
\EQ
\langle h_{\perp 3},(h_{\perp 1},h_{\perp 2})_\perp\rangle +\text{cyclic}
=0
\endEQ
from \eqref{adcyclictr}. 
Write $A(h_\perp) = \Dinvx[\u,h_\perp]_\parallel$. 
Then in the l.h.s of \eqref{sympeq}
the middle term in \eqref{varK} is given by:
\EQ
-\langle h_{\perp 3},(h_{\perp 1},A(h_{\perp 2}))\rangle
+\text{cyclic}
= 
\langle A(h_{\perp 2}),(h_{\perp 1},h_{\perp 3})_\parallel\rangle
+\text{cyclic}
\label{1stAterm}
\endEQ
from \eqref{adskewtr}. 
For the last term in \eqref{varK}, the l.h.s of \eqref{sympeq} simplifies:
\EQs
&&
-\langle h_{\perp 3},[\u,\Dinvx(h_{\perp 1},h_{\perp 2})_\parallel]\rangle
+\text{cyclic}
= 
\langle [\u,h_{\perp 3}],\Dinvx(h_{\perp 1},h_{\perp 2})_\parallel\rangle
+\text{cyclic}
\nonumber\\
&&
\equiv 
-\langle A(h_{\perp 3}),(h_{\perp 1},h_{\perp 2})\rangle +\text{cyclic}
= 
-\langle A(h_{\perp 2}),(h_{\perp 3},h_{\perp 1})\rangle +\text{cyclic}
\label{2ndAterm}
\endEQs
by cyclic rearrangement. 
Note $A(h_\perp)_\perp=0$. 
Hence, from \eqref{adsymm},
the terms \eqrefs{1stAterm}{2ndAterm} cancel, 
\EQ
\equiv  
\langle A(h_{\perp 2}),
(h_{\perp 1},h_{\perp 3})_\parallel - (h_{\perp 3},h_{\perp 1})_\parallel
\rangle =0
\endEQ
which completes the proof.

\subsection{Proof of compatibility of $\Hop,\Jop$}

Hamiltonian cosymplectic and symplectic operators $\Hop,\Jop$
are said to be {\it compatible (\ie/ comprise a Hamiltonian pair)} if
every linear combination $c_1\Hop +c_2\Jop\inv$ is a Hamiltonian operator. 
Equivalently \cite{Olver,Dorfman}, 
compatibility holds whenever 
$\Hop$ and $\Jop\inv$ have a vanishing Schouten bracket. 

The latter condition can be formulated using the same calculus of
vertical multi-vectors as employed in the proof of Proposition~4.8:
\EQ
\langle\uni\wedge( 
\uniopvar{\Hop}\Jop\inv(\uni) + \uniopvar{\Jop\inv}\Hop(\uni) )\rangle 
\equiv 0
\text{ modulo a total $x$-derivative }
\label{HJinveq}
\endEQ
where $\uni$ denotes the vertical uni-vector dual to $d\u$
in the $x$-jet space $J^\infty$ of $\u$. 
It is possible to simplify the l.h.s. of \eqref{HJinveq} by noting 
$\uniopvar{\Hop}\Jop\inv = -\Jop\inv(\uniopvar{\Hop}\Jop)\Jop\inv$
and then replacing $\uni$ by $\Jop(\uni)$ followed by using
the skew-adjoint property of $\Jop$ to get
\EQ
\langle
\uni\wedge \delta_{\Hop(\Jop(\uni))}\Jop(\uni) 
+ \Jop(\uni)\wedge(\univar\Hop)\Jop(\uni) 
\rangle 
\equiv 0 . 
\label{paireq}
\endEQ
Now, as the uni-vector defined by $\Jop(\uni)$ obviously enjoys 
the same formal properties as $\uni$ itself, 
this establishes that \eqref{paireq} is equivalent to 
condition \eqref{HJinveq}. 

{\bf Lemma~4.11: }{\it
The Schouten bracket of $\Hop,\Jop$ vanishes iff \eqref{paireq} holds. 
}

For verifying if $\Hop,\Jop$ satisfy \eqref{paireq}, 
the earlier multi-vector identities \eqsref{symmcomm}{utr} 
will be useful, in addition to the following identities:
\EQs
&&
\langle A(\uni), (B(\uni),C(\uni))_\perp\rangle + \text{cyclic}
=0 , 
\label{cyclicuni}\\
&&
(A(\uni),B(\uni))_\parallel=-(B(\uni),A(\uni))_\parallel , 
\label{skewuni}
\endEQs
and in particular
\EQ
(A(\uni),A(\uni))_\parallel=0 , 
\label{symmuni}
\endEQ
for linear operators $A,B,C$ from $\h_\perp$ into $\h$. 
Here, $(\cdot,\cdot) = [\ad(\ehook{x})\cdot,\cdot]$ as before. 

{\bf Theorem~4.12: }{\it
The cosymplectic and symplectic operators $\Hop,\Jop$ 
in Theorem~4.5 have vanishing Schouten bracket and hence comprise 
a Hamiltonian pair. 
}

Proof. 
It will be useful to introduce 
the uni-vectors 
$\aduni := \ad(\ehook{x})\inv\uni$
and 
$\adsquni := \ad(\ehook{x})\inv\aduni$,
with values in $\m_\perp$ and $\h_\perp$, 
having the same formal properties as the $\h_\perp$-valued uni-vector $\uni$. 
(In particular, $\uni=\ad(\ehook{x})\aduni = \ad(\ehook{x})^2\adsquni$.)
To begin, 
note 
$\langle\uni\wedge \delta_{\Hop(\Jop(\uni))}\Jop(\uni) \rangle
= \langle\aduni\wedge \delta_{\Hop(\Jop(\uni))}\Kop(\aduni) \rangle$
due to $\ad(\ehook{x})$ being skew-adjoint 
with respect to the Cartan-Killing inner product. 
Now consider
\EQ\label{HJvarK}
\delta_{\Hop(\Jop(\uni))}\Kop(\aduni) 
= 
-[\Hop(\Jop(\uni)),\Dinvx[\u,\aduni]_\parallel]
- [\u,\Dinvx[\Hop(\Jop(\uni)),\aduni]_\parallel]
+[\Hop(\Jop(\uni)),\aduni]_\perp
\endEQ
and write $A(\aduni):= \Dinvx[\u,\aduni]_\parallel$. 
In the l.h.s. of \eqref{paireq},
the first term in \eqref{HJvarK} yields
\EQ
-\langle\aduni\wedge[\Hop(\Jop(\uni)),A(\aduni)] \rangle
= -\langle A(\aduni) \wedge[\aduni,\Hop(\Jop(\uni))] \rangle
\endEQ
and the second term in \eqref{HJvarK} simplifies:
\EQ
-\langle\aduni\wedge[\u,\Dinvx[\Hop(\Jop(\uni)),\aduni]_\parallel] \rangle
\equiv -\langle A(\aduni) \wedge[\aduni,\Hop(\Jop(\uni))] \rangle .
\endEQ
Hence these two terms add to give, from \eqref{HJops},
\EQ\label{varHJterms1}
-2\langle A(\aduni) \wedge[\aduni,\Dx\Jop(\uni)] \rangle
-2\langle A(\aduni) \wedge[[\u,\Jop(\uni)]_\perp,\aduni] \rangle
+ 2\langle A(\aduni) \wedge[[\u,B(\uni)],\aduni] \rangle
\endEQ
where $B(\aduni):=\Dinvx[\u,\Jop(\uni)]_\parallel$. 
Next, in the l.h.s. of \eqref{paireq}, 
the third term in \eqref{HJvarK} expands out to yield
\EQ\label{varHJterms2}
\langle \aduni\wedge [\Hop(\Jop(\uni)),\aduni] \rangle
= \langle [\aduni,\aduni] \wedge \Dx\Jop(\uni) \rangle
+ \langle [\aduni,\aduni] \wedge [\u,\Jop(\uni)]_\perp \rangle
-\langle [\aduni,\aduni] \wedge [\u,B(\uni)] \rangle . 
\endEQ
Note the six terms given by \eqrefs{varHJterms1}{varHJterms2} 
are linear in $\J(\uni)$. 
Six similar terms come from 
$\langle\Jop(\uni)\wedge (\univar\Hop)\Jop(\uni) \rangle$
as follows. 
First consider
\EQ\label{varH}
(\univar\Hop)\Jop(\uni)
= 
-[\uni,\Dinvx[\u,\Jop(\uni)]_\parallel]
- [\u,\Dinvx[\uni,\Jop(\uni)]_\parallel]
+[\uni,\Jop(\uni)]_\perp . 
\endEQ
In the l.h.s. of \eqref{paireq}, 
the second term in \eqref{varH} simplifies via \eqref{cyclictr}
\EQ
-\langle \Jop(\uni)\wedge [\u,\Dinvx[\uni,\Jop(\uni)]_\parallel]\rangle
\equiv -\langle B(\uni)\wedge [\uni,\Jop(\uni)]\rangle
\endEQ
which then adds with the first term in \eqref{varH}, 
yielding by \eqrefs{skewuni}{HJops},
\EQ
-2\langle B(\uni)\wedge [\uni,\Jop(\uni)]\rangle
= -2\langle B(\uni)\wedge (\aduni,\Jop(\uni))_\parallel\rangle
= -2\langle B(\uni)\wedge [\Kop(\aduni),\aduni]\rangle . 
\endEQ
This expands out to give three terms:
first
\EQ\label{varHterm1}
-2\langle B(\uni)\wedge [\Dx\aduni,\aduni]\rangle
\equiv 
\langle [\u,\Jop(\uni)]_\parallel\wedge [\aduni,\aduni]\rangle
\endEQ
by \eqref{symmcomm};
next
\EQ\label{varHterm2}
-2\langle B(\uni)\wedge [[\u,\aduni]_\perp,\aduni]\rangle
= -2\langle B(\uni)\wedge [[\u,\aduni],\aduni]_\parallel\rangle
= -\langle B(\uni)\wedge [\u,[\aduni,\aduni]]\rangle
\endEQ
by $[[\u,\uni]_\parallel,\uni]_\parallel=0$
from the Lie bracket relations \eqref{commpar}
followed by \eqref{unibrac};
third
\EQ\label{varHterm3}
2\langle B(\uni)\wedge [[\u,A(\aduni)],\aduni]\rangle . 
\endEQ
The last term remaining in \eqref{varH} can be rearranged
in the l.h.s. of \eqref{paireq}:
\EQs
&&
\langle \Jop(\uni)\wedge [\uni,\Jop(\uni)]\rangle
= \langle \Jop(\uni)\wedge (\aduni,\Jop(\uni))_\perp\rangle
\nonumber\\
&&
= - \langle \aduni\wedge (\Jop(\uni),\Jop(\uni))_\perp\rangle
- \langle \Jop(\uni)\wedge (\Jop(\uni),\aduni)_\perp\rangle
= 2 \langle \Jop(\uni)\wedge [\aduni,\Kop(\aduni)]\rangle
\endEQs
via \eqrefs{cyclicuni}{cyclictr}. 
This term expands out to give three terms:
\EQ\label{varHterms}
\langle \Jop(\uni)\wedge \Dx[\aduni,\aduni]\rangle
-2\langle \Jop(\uni)\wedge [\aduni,[\u,A(\aduni)]]\rangle
+2\langle \Jop(\uni)\wedge [\aduni,[\u,\aduni]_\perp]\rangle
\endEQ
via \eqref{symmcomm}. 

Now the term \eqref{varHterm2} cancels the last term in \eqref{varHJterms2}
by \eqref{utr},
while the first terms in \eqrefs{varHterms}{varHJterms2}
reduce to $\Dx\langle\Jop(\uni)\wedge[\aduni,\aduni]\rangle \equiv 0$.

The remaining terms in \eqrefs{varHterms}{varHJterms2},
along with the three terms in \eqref{varHJterms1}
and the two terms given by \eqrefs{varHterm1}{varHterm3},
constitute a cubic polynomial in $\u$ 
with coefficients linear in $\Jop(\uni)$.
The linear $\u$-terms in this polynomial consist of:
\EQ\label{lin1}
2\langle \Jop(\uni)\wedge [\aduni,[\u,\aduni]_\perp]\rangle
= \langle \Jop(\uni)\wedge [\u,[\aduni,\aduni]]\rangle
-2\langle \Dx A(\aduni)\wedge [\aduni,\Jop(\uni)]\rangle
\endEQ
via \eqrefs{unibrac}{cyclictr};
plus
\EQ\label{lin2}
\langle [\aduni,\aduni]\wedge[\u,\Jop(\uni)]\rangle 
\endEQ
and
\EQ\label{lin3}
-2\langle A(\aduni)\wedge[\aduni,\Dx\Jop(\uni)]\rangle . 
\endEQ
When combined, these four terms reduce to 
\EQ\label{linu}
\equiv 2\langle A(\aduni)\wedge[\Dx\aduni,\Jop(\uni)]\rangle . 
\endEQ
Next, the quadratic $\u$-terms in the polynomial yield
\EQ\label{quadu}
-2\langle \Jop(\uni)\wedge[\aduni,[\u,A(\aduni)]]\rangle 
-2\langle A(\aduni)\wedge[\aduni,[\u,\Jop(\uni)]_\perp]\rangle 
=-2\langle \aduni\wedge[\u,[A(\aduni),\Jop(\uni)]]\rangle 
\endEQ
via \eqrefs{cyclictr}{unijacobi},
with $[\aduni,[\u,\Jop(\uni)]_\parallel]_\parallel=0$
by the Lie bracket relation \eqref{commpar}. 
Similarly, the cubic terms combine to produce
\EQs
&&
2\langle [[\u,A(\aduni)],\uni]\wedge B(\aduni)\rangle 
+2\langle A(\aduni)\wedge [[\u,B(\aduni)],\aduni]\rangle 
= 2\langle \aduni\wedge[\u,[A(\aduni),B(\uni)]\rangle 
\nonumber\\
&&
= -2\langle \Dx A(\aduni)\wedge[A(\aduni),B(\aduni)]\rangle 
\equiv \langle [A(\aduni),A(\aduni)]\wedge[\u,\Jop(\uni)]\rangle 
\label{cubu}
\endEQs
where the middle step uses
$[A(\aduni),B(\aduni)]_\perp=0$ 
due to the Lie bracket relations \eqref{commperp}. 

To proceed, we make use of the identity 
\EQ\label{Jid}
\Jop(\uni) 
=-\ad(\ehook{x})\inv\Kop(\aduni)
= -\Dx\adsquni -C(\aduni) +[\adu,A(\aduni)]
\endEQ
with 
$C(\aduni):=\ad(\ehook{x})\inv[\u,\aduni]_\perp$
and 
$\adu := \ad(\ehook{x})\inv\u$
where the last term in \eqref{Jid} arises from 
$[\u,A(\aduni)]=\ad(\ehook{x})[\adu,A(\aduni)]$
as $\ad(\ehook{x})$ annihilates $A(\aduni)$. 
The terms \eqref{linu}, \eqref{quadu}, \eqref{cubu} 
expand out to become a quintic polynomial in $\u$ as follows. 
There is a single linear term, coming from \eqref{linu}:
\EQ
-2\langle A(\aduni)\wedge[\Dx\aduni,\Dx\adsquni]\rangle 
=-2\langle A(\aduni)\wedge(\Dx\adsquni,\Dx\adsquni)_\parallel\rangle 
=0
\endEQ
by \eqref{symmuni}. 
Next, one quadratic term comes from \eqref{linu}:
\EQs
&&
-2\langle A(\aduni)\wedge[\Dx\aduni,C(\aduni)]\rangle 
= -2\langle A(\aduni)\wedge(\Dx\adsquni,C(\aduni))_\parallel\rangle 
= 2\langle A(\aduni)\wedge[[\u,\aduni]_\perp,\Dx\adsquni]\rangle 
\nonumber\\
&&
= 2\langle \Dx\adsquni\wedge[A(\aduni),[\u,\aduni]]\rangle 
\label{quadterm}
\endEQs
via \eqrefs{symmuni}{cyclictr} combined with 
$[A(\aduni),[\u,\aduni]_\parallel]_\perp=0$
by the Lie bracket relations \eqref{commperp}. 
This term \eqref{quadterm} now cancels 
the quadratic term given by \eqref{quadu}, 
\EQ
2\langle \aduni\wedge[\u,[A(\aduni),\Dx\adsquni]]\rangle 
= -2\langle [\u,\aduni]\wedge[A(\aduni),\Dx\adsquni]\rangle 
\endEQ
through \eqref{cyclictr}. 
We then have only cubic and higher degree terms remaining to consider. 
One cubic term comes from \eqref{quadu}:
\EQ
2\langle \aduni\wedge[\u,[A(\aduni),C(\aduni)]]\rangle 
= -2\langle A(\aduni)\wedge[[\u,\aduni],C(\aduni)]\rangle 
= -2\langle A(\aduni)\wedge(C(\aduni),C(\aduni))_\parallel\rangle 
=0
\endEQ
by \eqref{symmuni},
with $[[\u,\aduni]_\parallel,C(\aduni)]_\parallel=0$
by the Lie bracket relations \eqref{commpar}. 
Another cubic term arises from \eqref{linu}:
\EQ\label{cubterm1}
2\langle A(\aduni)\wedge[\Dx\aduni,[\adu,A(\aduni)]]\rangle 
= \langle \Dx\aduni\wedge[\adu,[A(\aduni),A(\aduni)]]\rangle 
= \langle [A(\aduni),A(\aduni)]\wedge(\Dx\adsquni,\adu)_\parallel\rangle 
\endEQ
via \eqrefs{cyclictr}{unibrac}. 
A similar cubic term is given by \eqref{cubu}:
\EQ\label{cubterm2}
\langle [A(\aduni),A(\aduni)]\wedge[\u,\Dx\adsquni]\rangle 
= - \langle [A(\aduni),A(\aduni)]\wedge(\adu,\Dx\adsquni)_\parallel\rangle . 
\endEQ
Then \eqrefs{cubterm1}{cubterm2} cancel by \eqref{adsymm}. 
There is a similar cancellation of quartic terms coming from 
\eqrefs{quadu}{cubu}:
\EQ
-2\langle \aduni\wedge[\u,[A(\aduni),[\adu,A(\aduni)]]]\rangle 
= \langle [A(\aduni),A(\aduni)]\wedge[[\u,\aduni]_\perp,\adu]\rangle 
= \langle [A(\aduni),A(\aduni)]\wedge(C(\aduni),\adu)_\parallel\rangle 
\endEQ
and
\EQ
-\langle [A(\aduni),A(\aduni)]\wedge[\u,C(\aduni)]\rangle 
= -\langle [A(\aduni),A(\aduni)]\wedge(\adu,C(\aduni))_\parallel\rangle .
\endEQ
As a result, the only term remaining is quintic,
which comes from \eqref{cubu}:
\EQ\label{quinterm}
\langle [A(\aduni),A(\aduni)]\wedge[\u,[\adu,A(\aduni)]]\rangle . 
\endEQ
We now employ the Jacobi identity combined with the identity
\EQ
[\u,[\adu,A(\aduni)]]_\parallel
= (\adu,[\adu,A(\aduni)])_\parallel
= ([\adu,A(\aduni)],\adu)_\parallel
= [[\u,A(\aduni)],\adu]_\parallel
\endEQ
to get
\EQ
[\u,[\adu,A(\aduni)]]_\parallel
= \frac{1}{2}[[\u,\adu],A(\aduni)]_\parallel . 
\endEQ
Hence \eqref{quinterm} simplifies to
\EQ
\frac{1}{2}\langle [A(\aduni),A(\aduni)]\wedge[[\u,\adu],A(\aduni)]\rangle 
= \frac{1}{2}\langle [\u,\adu]\wedge[[A(\aduni),A(\aduni)],A(\aduni)]\rangle 
=0
\endEQ
by \eqref{jacobicomm}. 
Consequently, 
all terms in the l.h.s. of \eqref{paireq} have been shown to vanish
modulo a total $x$-derivative, 
thereby completing the proof.

\section{Bi-Hamiltonian hierarchies of non-stretching curve flows and soliton equations}

In any Riemannian symmetric space $M=G/H$, 
the frame structure equations of non-stretching curve flows $\map(t,x)$
geometrically encode a group-invariant bi-Hamiltonian structure
for the induced flow on the components of the principal normal vector
along the curve. 
This structure looks simplest as stated in Theorem~4.5
using an $H$-parallel moving frame formulation (cf Definition~3.3)
expressed in terms of a $\m$-valued linear coframe $\e$
and its associated $\h$-valued linear connection $\conx$,
obeying properties (i) to (iv) summarized in Lemma~4.1.
Such a framing of $\map(t,x)$ arises directly from 
the $\g$-valued zero-curvature Cartan connection $\gconx=\e+\conx$
in the Klein geometry associated with $G$ viewed as an $H$-bundle over $M$,
where $\g=\m\oplus\h$ is the corresponding Lie algebra decomposition of $G$
and $\m\simeq_G T_x M$. 
In particular, $\e$ and $\conx$ provide a soldering of 
the Klein geometry of $G/H$ onto the Riemannian geometry of $M$,
such that the frame structure equations coincide with 
the Cartan equations for torsion and curvature 
(cf Theorem~2.3 and Corollary~2.4) 
of the Riemannian connection $\covder{t},\covder{x}$ 
on the two-dimensional surface of the flow in $M$. 

The bi-Hamiltonian structure consists of compatible Hamiltonian
(cosymplectic and symplectic) operators $\Hop$ and $\Jop$,
which yield a (hereditary) recursion operator 
\EQs
\Rop &=& \Hop\Jop =
-(\Kop\ad(\ehook{x})\inv)^2
\nonumber\\
&=&
-( \ad(\ehook{x})\inv\Dx +[\u,\ad(\ehook{x})\inv\cdot\ ]_\perp
- [\u,\Dinvx[\u,\ad(\ehook{x})\inv\cdot\ ]_\parallel] )^2
\label{recursop}
\endEQs
with respect to the flow variable $\u$. 
Due to the $H$-parallel property of $\e$ and $\conx$,
$\e_x=\e\hook\mapder{x}$ is a fixed unit vector in $\m$
representing the moving frame components of the tangent vector 
$\mapder{x}$ along the curve,
while $\u=\conx_x=\conx\hook\mapder{x}$ is related to 
the moving frame components of the principal normal vector 
$\covder{x}\mapder{x}$ along the curve
through the corresponding $\m_\perp$-valued expression
$[\u,\e_x]=-\ad(\e_x)\u= \e\hook\covder{x}\mapder{x}$. 
Note $\ad(\ehook{x})\inv$ gives an isomorphism between the perp spaces 
$\h_\perp$ and $\m_\perp$ of the centralizer subspaces 
$\h_\parallel$ and $\m_\parallel$ with respect to $\e_x$ in $\m\oplus\h=\g$,
with $\ad(\e_x)\h_\parallel= \ad(\e_x)\m_\parallel=0$. 
Hence the linear operator $\Kop\ad(\ehook{x})\inv$ maps
$\h_\perp$-valued functions into $\m_\perp$-valued functions,
and vice versa. 
Thus the adjoint of the recursion operator is given by 
\EQs
\adRop &=& \Jop\Hop = 
-(\ad(\ehook{x})\inv\Kop)^2
\nonumber\\
&=&
-( \ad(\ehook{x})\inv\Dx +\ad(\ehook{x})\inv[\u,\cdot\ ]_\perp
- \ad(\ehook{x})\inv[\u,\Dinvx[\u,\cdot\ ]_\parallel] )^2 . 
\endEQs

On the $x$-jet space 
$J^\infty=(x,\u,\u_x,\u_{xx},\ldots)$ of the flow variable $\u$,
the $x$-translation vector field $\partial/\partial x$ 
is an obvious symmetry of the operators $\Hop,\Jop$ 
and hence of the recursion operator $\Rop$.
As a consequence, 
from general results due to Magri \cite{Magri1,Magri2}
on Hamiltonian recursion operators \cite{Olver}, 
$\Rop$ will generate a hierarchy of commuting Hamiltonian vector fields
with respect to the Poisson bracket, 
where the hierarchy starts with 
the evolutionary form of the vector field $\partial/\partial x$. 
Moreover, an associated hierarchy of involutive covector fields
arises from the canonical pairing provided by the symplectic $2$-form.

{\bf Theorem~5.1: }{\it
A commuting hierarchy of Hamiltonian vector fields 
$h^\perp_\nth{n}\cdot\partial/\partial\u$
is given by the $\h_\perp$-valued functions 
\EQ\label{hperphierarchy}
h^\perp_\nth{n}:=\Rop^n(\u_x) ,\quad
(n=0,1,2,\ldots)
\endEQ
satisfying the Poisson bracket relation
\EQ
\delta_{h^\perp\nth{n}}\ \cdot=\{\cdot,\ham{H}\supth{n}\}_\Hop
\endEQ
for some Hamiltonian functionals
$\ham{H}\supth{n}=\int\Ham\supth{n}(x,\u,\u_x,\ldots)$,
with respect to the Hamiltonian operator $\Hop$. 
The variational derivative of each such functional yields
a dual vector field 
$\varpi^\perp_\nth{n}\cdot d\u$ given by 
the $\h_\perp$-valued functions 
\EQ\label{conxhierarchy}
\varpi^\perp_\nth{n}:=\delta\Ham\supth{n}/\delta\u= \adRop^n(\u) ,\quad
(n=0,1,2,\ldots)
\endEQ
forming a hierarchy of involutive variational covector fields. 
In particular, at the bottom of these hierarchies, 
$\Ham\supth{0}=-\frac{1}{2}\langle\u,\u\rangle$ 
is the Hamiltonian for the vector field given by 
$h^\perp_\nth{0} =\u_x$
corresponding to the generator of $x$-translations in evolutionary form,
whose dual covector field is given by 
$\varpi^\perp_\nth{0}=\u$. 
}

{\bf Corollary~5.2: }{\it
These hierarchies are related by the Hamiltonian operators
\EQ\label{hamstructure}
\varpi^\perp_\nth{n+1}=\Jop(h^\perp_\nth{n}) ,\quad
h^\perp_\nth{n} =\Hop(\varpi^\perp_\nth{n})
\endEQ
with 
$\langle\varpi^\perp_\nth{n},h^\perp_\nth{n}\rangle \equiv 0$
modulo a total $x$-derivative, 
since $\Hop,\Jop$ are skew-adjoint. 
}

The form of these operators under a scaling of the variables $x$ and $\u$
shows that both hierarchies possess the mKdV scaling symmetry
$x\rightarrow \lambda x$, $\u\rightarrow \lambda\inv \u$. 
Thereby 
$h^\perp_\nth{n}$ and $\Ham\supth{n}$ each have scaling weight $2+2n$,
and $\varpi^\perp_\nth{n}$ hence has scaling weight $1+2n$.
From variational scaling methods in \cite{Anco2}, 
it follows that $\Ham\supth{n}$ can be expressed explicitly in terms of
$h^\perp_\nth{n}$ or $\varpi^\perp_\nth{n}$: 
\EQ\label{Hform}
(1+2n)\Ham\supth{n} 
=-\Dinvx\langle\u,h^\perp_\nth{n}\rangle
=-\Dinvx\langle\u,\Hop(\varpi^\perp_\nth{n})\rangle . 
\endEQ
The derivation of this formula is fairly simple. 
Let $\delta_{\rm s}\u= -x\u_x-\u =-\Dx(x\u)$
and $\delta_{\rm s}\Ham= -x\Dx\Ham-p\Ham$
be the evolutionary form of the mKdV scaling symmetry generator
on $\u$ and $\Ham(x,\u,\u_x,\ldots)$,
with respective scaling weights $1$ and $p\neq 1$.
Then we have
\EQ
\delta_{\rm s}\Ham\equiv (1-p)\Ham
\endEQ
modulo a total $x$-derivative.
On the other hand, 
\EQ
\delta_{\rm s}\Ham 
\equiv -\langle\delta_{\rm s}\u,\delta\Ham/\delta\u\rangle
= \langle\Dx(x\u),\varpi^\perp\rangle
\equiv -x\langle\u,\Dx\varpi^\perp\rangle
= -x\langle\u,\Hop(\varpi^\perp)\rangle , 
\endEQ
where the last line uses the identity
$\langle\u,[\u,A(\varpi^\perp)]\rangle 
= \langle A(\varpi^\perp),[\u,\u]\rangle =0$
holding for any linear operator $A$. 
Finally, the identity 
$\langle\u,\Hop(\varpi^\perp)\rangle
= \Dx\Dinvx\langle\u,\Hop(\varpi^\perp)\rangle$
yields
$\delta_{\rm s}\Ham 
\equiv \Dinvx\langle\u,\Hop(\varpi^\perp)\rangle$
modulo a total $x$-derivative,
from which we obtain 
$\Ham=(1-p)\inv\Dinvx\langle\u,\Hop(\varpi^\perp)\rangle$.

\subsection{Group-invariant bi-Hamiltonian soliton equations}

The entire hierarchy of vector fields \eqref{hperphierarchy}
has two compatible Hamiltonian structures $\Hop$ and $\Jop\inv$:
\EQ\label{bihamhperp}
h^\perp_\nth{n}
=\Hop( \delta\Ham\supth{n}/\delta\u )
=\Jop\inv( \delta\Ham\supth{n+1}/\delta\u ) ,\quad
(n=0,1,2,\ldots) . 
\endEQ
The operator $\Eop:=\Rop\Hop$ provides an alternate
Hamiltonian structure 
\EQ
h^\perp_\nth{n}
=\Eop( \delta\Ham\supth{n-1}/\delta\u ) \quad\text{ (for $n>0$), }
\endEQ
with $\Hop$ and $\Eop$ comprising an explicit Hamiltonian pair
(of cosymplectic operators).
As a main result, this bi-Hamiltonian structure \eqref{bihamhperp}
produces a hierarchy of bi-Hamiltonian flows on $\u$
as given by linear combinations 
$h^\perp = h^\perp_\nth{n+1} + h^\perp_\nth{n}$
of the vector fields \eqref{hperphierarchy}
according to the form of the flow equation \eqref{ufloweq}. 

{\bf Theorem~5.3: }{\it
The flow equation \eqref{ufloweq} on $\u$ yields a hierarchy of
bi-Hamiltonian evolution equations
\EQ\label{evolhierarchy}
\u_t =\Hop( \delta\Ham\supth{n,1}/\delta\u )
=\Jop\inv( \delta\Ham\supth{n+1,1}/\delta\u ) 
\endEQ
(hereafter called the $+1+n$ flow)
for $n=0,1,2,\ldots$,
where 
$\Ham\supth{n,1}:=\Ham\supth{n}+ \Ham\supth{n-1}$.
Each of these $\h_\perp$-valued evolution equations is 
invariant under the equivalence group $\equivH \subset H^*$
consisting of rigid ($x$-independent) linear transformations
$\Ad(h\inv)$ that preserve $\e_x$. 
}

These evolution equations \eqref{evolhierarchy} 
concretely describe 
group-invariant multicomponent soliton equations
when $\u$ is expressed by real-valued components
in an expansion with respect to any fixed basis \eqrefs{onbasis}{commbasis}
for the vector space 
$\h_\perp\simeq\m_\perp$. 
These components will naturally divide into irreducible representations
$\u^\alpha$ under the action of the group $\equivH$,
corresponding to a direct sum decomposition 
$\h_\perp = \oplus_\alpha \h_\perp^\alpha$
where each subspace $\h_\perp^\alpha$ is an eigenspace of $\Ad(\equivH)$. 
The specific algebraic structure of these eigenspaces 
(which will depend on the structure of the group $\equivH$)
defines the algebraic content of the variables $\u^\alpha$: \eg/
scalars, real vectors, complex vectors or vector pairs, 
scalar-vector pairs, etc..
For instance, in the example $M=SO(N+1)/SO(N)$
studied in \cite{sigmapaper}, 
the Lie algebra of $\equivH$ is 
$\h_\parallel=\alg{so}(N-1) \subset \h=\alg{so}(N)$, 
whence $\h_\perp \simeq \Rnum^{N-1} \simeq \m_\perp$
with this vector space being irreducible (\ie/ $\alpha=1$). 
Thus, in this case, 
$\u\in\Rnum^{N-1}$ is algebraically identified 
as a single real vector variable,
and the hierarchy of vector evolution equations \eqref{evolhierarchy}
describes $SO(N-1)$-invariant multicomponent soliton equations
for $\u=(\u^1,\ldots,\u^{N-1})$. 

In general the bi-Hamiltonian evolution equations \eqref{evolhierarchy}
can be rewritten in terms of $\m_\perp$-valued variables
more geometrically related to the flow $\map(t,x)$ as follows:
\EQ\label{princnor}
N=\covder{x}\mapder{x} 
\leftrightarrow \nu:=-\ad(\e_x)\u
\endEQ
represents the principal normal vector along the curve; 
\EQ\label{perpparflows}
(\mapder{t})_\perp
\leftrightarrow h_\perp:=\ad(\e_x)\inv h^\perp ,\quad
(\mapder{t})_\parallel
\leftrightarrow h_\parallel
\endEQ
represent the algebraically perpendicular and parallel parts 
of the flow vector;
\EQ\label{flowprincnor}
\covder{t}\mapder{x} 
\leftrightarrow \varpi_\perp:=-\ad(\e_x)\varpi^\perp 
= -\ad(\e_x)(\varpi^\perp + \varpi^\parallel)
\endEQ
represents the principal normal vector relative to the flow. 
These geometrical variables inherit a natural Hamiltonian 
cosymplectic and symplectic structure from the results
in Theorem~5.1 and Corollary~5.2
expressed with respect to the flow variable
\EQ\label{vurel}
\v:= \ad(\e_x)\inv\u = \adsq\inv(\nu) ,\quad
\adsq=-\ad(\e_x)^2
\endEQ
in the following way. 
To begin, the induced flow equation takes the form 
\EQ\label{vteq}
\v_t =\tilde\Hop(\varpi_\perp) + h_\perp ,\quad
\varpi_\perp = \tilde\Jop(h_\perp) 
\endEQ
where
\EQ\label{tildeHJops}
\tilde\Hop = -\ad(\e_x)\inv\Kop|_{\h_\perp}\ad(\e_x)\inv , \quad
\tilde\Jop = \Kop|_{\m_\perp}
\endEQ
are compatible Hamiltonian cosymplectic and symplectic operators
given in terms of
\EQ
\Kop =
\Dx +[\ad(\e_x)\v,\cdot\ ]_\perp 
- [\ad(\e_x)\v,\Dinvx[\ad(\e_x)\v,\cdot\ ]_\parallel]_\perp
\endEQ
from \eqrefs{Kop}{vurel}. 

{\bf Proposition~5.4: }{
On the $x$-jet space of geometrical flows with respect to $\v$,
$h_\perp\cdot\partial/\partial\v$ is a Hamiltonian vector field
\EQ
\delta_{h^\perp}(\cdot)
= \{\cdot, \ham{H}\}_{\tilde\Hop}
= -\int\langle\delta(\cdot)/\delta\v,
\tilde\Hop(\delta\ham{H}/\delta\v)\rangle dx
\endEQ
with respect to the Poisson bracket 
$\{\cdot,\cdot\}_{\tilde\Hop}$,
and $\varpi_\perp\cdot d\v$ is a variational covector field
\EQ
\varpi_\perp = \delta\ham{F}/\delta\v
\endEQ
satisfying the duality relation 
\EQ
(\varpi_\perp\cdot d\v)\hook\cdot\ 
= \sympform(\cdot\ ,h_\perp\cdot\partial/\partial\v)_{\tilde\Jop}
= -\int\langle\cdot\ ,\tilde\Jop(h_\perp)\rangle dx
\endEQ
with respect to the symplectic $2$-form 
$\sympform(\cdot,\cdot)_{\tilde\Jop}$. 
}

{\bf Theorem~5.5: }{\it
There is a bi-Hamiltonian hierarchy of evolution equations on
the principal normal components \eqref{princnor} 
in a $H$-parallel moving frame, represented by 
\EQ\label{vevolhierarchy}
\v_t  
=\tilde\Hop( \delta\tilde\Ham\supth{n,1}/\delta\v )
=\tilde\Jop\inv( \delta\tilde\Ham\supth{n+1,1}/\delta\v ) ,\quad
(n=0,1,2,\ldots)
\endEQ
with Hamiltonians 
$\tilde\Ham\supth{n,1}:=\Ham\supth{n}+ \Ham\supth{n-1}$
expressed in terms of the $\m_\perp$-valued variables
$(\v,\v_x,\ldots)$ via \eqrefs{vurel}{Hform}. 
Correspondingly, the normal flow components
\EQ\label{vhperphierarchy}
h_\perp^\nth{n} = \tilde\Rop{}^n(\v_x) 
=\tilde\Hop(\varpi_\perp^\nth{n}) ,\quad
(n=0,1,2,\ldots)
\endEQ
and the principal normal components with respect to the normal flow 
\EQ\label{vwhierarchy}
\varpi_\perp^\nth{n} = \tilde\adRop{}^{n}(\adsq(\v)) 
=\tilde\Jop(h_\perp^\nth{n-1}) ,\quad
(n=1,2,\ldots)
\endEQ
represent hierarchies of 
commuting Hamiltonian vector fields 
$h_\perp^\nth{n}\cdot\partial/\partial\v$
and involutive variational covector fields
$\varpi_\perp^\nth{n}\cdot d\v$
generated by the recursion operator 
\EQ
\tilde\Rop=\tilde\Hop\tilde\Jop
= -(\ad(\e_x)\inv\Kop)^2
\endEQ
and its adjoint 
\EQ
\tilde\adRop=\tilde\Jop\tilde\Hop
= -(\Kop\ad(\e_x)\inv)^2 . 
\endEQ
}

This version of the flow equation and its bi-Hamiltonian structure
is closest to the notation used in \cite{sigmapaper,imapaper}.
Note the invariance group of the operators $\tilde\Hop$ and $\tilde\Jop$
is $H^*_\parallel \subset H^*\simeq\Ad(H)$,
and the scaling weights of $h_\perp^\nth{n}$ and $\varpi_\perp^\nth{n}$
are $2+2n$ and $1+2n$
under the mKdV scaling group 
$x\rightarrow \lambda x$, $\v\rightarrow \lambda\inv\v$. 

Through the geometric correspondences \eqsref{princnor}{flowprincnor},
the evolution equations \eqref{vevolhierarchy} produce 
a bi-Hamiltonian hierarchy of non-stretching curve flows in $M=G/H$. 
The Hamiltonians are related to the tangential part of the flow
by 
\EQ
g(\mapder{t},\mapder{x}) 
\leftrightarrow 
-\langle h_\parallel\supth{n},\e_x\rangle
= (1+2n)\Ham\supth{n} 
\endEQ
arising from \eqrefs{hparwpareq}{Hform},
where
\EQ\label{vham}
h_\parallel =\Dinvx[\v,\ad(\e_x)h_\perp]_\parallel ,\quad
\Ham\supth{n} =- \frac{1}{1+2n}\Dinvx\langle\v,\adsq(h_\perp^\nth{n})\rangle .
\endEQ

{\it Remarks: }
\vskip0pt
(1) 
The infinitesimal invariance group of the hierarchy \eqref{vevolhierarchy} 
can be identified with the Lie subalgebra $\h_\parallel \subset \h$
given by the centralizer of $\e_x$ in the Lie algebra of the group $H$. 
\vskip0pt
(2) 
The bi-Hamiltonian structure \eqref{hamstructure} of the hierarchy 
depends on the constant unit vector $\e_x$ in $\m\simeq \g/\h$
determined by the framing of the curve flow $\map(t,x)$.
\vskip0pt
(3) 
Two curve flows encode equivalent (isomorphic) bi-Hamiltonian structures 
iff $\e_x$ lies on the same orbit of the linear isotropy group $H^*$ in $\m$
for both flows,
where all such orbits are parametrized by 
the unit-norm elements in the Weyl chamber
$a_*(\m)$ (cf Proposition~3.2)
contained in a fixed maximal abelian subspace $a\subset \m$.
\vskip0pt
(4) Without loss of generality, in a given equivalence class, 
a representative $e_x$ can be chosen such that the linear map 
$\ad(\e_x)^2$ is diagonal on $\m_\perp$. 

Consequently, 
every Riemannian symmetric space $M=G/H$ will possess
distinct bi-Hamiltonian hierarchies of 
non-stretching curve flows 
characterized by the distinct elements in a canonical Weyl chamber 
$a_*(\m)\subset\m\simeq \g/\h\simeq_G T_x M$. 
For the example $M=SO(N+1)/SO(N)$,
it is well-known that every maximal abelian subspace 
$a\subset \m = \alg{so}(N+1)/\alg{so}(N) \simeq \Rnum^{N}$
has $\dim a=1$,
whence the Weyl chamber $a_*(\m)$ has a single unit-norm element.
As a result, up to equivalence, 
there is a single bi-Hamiltonian hierarchy of curve flows
in $M=SO(N+1)/SO(N)$.

The equations of the curve flows on $\map(t,x)$ in a given hierarchy 
in $M=G/H$
can be readily derived in an explicit form from 
the geometric correspondences \eqsref{princnor}{flowprincnor}
by means of Proposition~4.4.

\subsection{mKdV flows}

The $+1$ flow on $\v$ will now be derived in detail.
This will require working out the first higher-order Hamiltonian structures
in the hierarchy \eqrefs{vhperphierarchy}{vwhierarchy}.
At the bottom of the hierarchy, note
\EQs
&&
h_\perp^\nth{0} =\v_x , 
\label{hperp0th}\\
&&
\Ham\supth{0}=-\langle h_\parallel^\nth{0},\e_x\rangle
= -\Dinvx\langle\v,\adsq(h_\perp^\nth{0})\rangle
=-\frac{1}{2}|\ad(\e_x)\v|^2 , 
\label{ham0th}\\
&&
\varpi_\perp^\nth{0} =\delta\Ham\supth{0}/\delta\v=\adsq(\v) , 
\label{w0th}
\endEQs
along with 
\EQ
h_\parallel^\nth{0} = \Dinvx[v,\ad(\e_x)h_\perp^\nth{0}]_\parallel
=-\frac{1}{2}[\ad(\e_x)\v,\v]_\parallel ,
\endEQ
where, recall, $\adsq=-\ad(\e_x)^2$. 
The next order in the hierarchy is given by 
\EQs
\delta\Ham\supth{1}/\delta\v
&=&
\varpi_\perp^\nth{1} 
=\tilde\Jop(h_\perp^\nth{0}) 
= \tilde\adRop(\adsq(\v)) 
\nonumber\\
&=&
\v_{xx} +[\ad(\e_x)\v,\v_x -\frac{1}{2} [\ad(\e_x)\v,\v]_\parallel] , 
\label{w1st}
\endEQs
\EQs
h_\perp^\nth{1} 
&=&
\tilde\Hop(\varpi_\perp^\nth{1})
= \tilde\Rop(\v_x)
\nonumber\\
&=&
\adsq\inv(\v_{xxx}) 
+\adsq\inv([\ad(\e_x)\v,\v_x-\frac{1}{2}[\ad(\e_x)\v,\v]_\parallel]_\perp)_x 
\nonumber\\&&\qquad
- \ad(\e_x)\inv[ \ad(\e_x)\v, 
\ad(\e_x)\inv(\v_{xx} +[\ad(\e_x)\v,\v_x]_\perp)
\nonumber\\&&\qquad
-\frac{1}{2}[\v,[\ad(\e_x)\v,\v]_\parallel] ]
-[\v,[\v,\v_x+\frac{1}{3}[\ad(\e_x)\v,\v]_\perp]_\parallel] , 
\label{hperp1st}
\endEQs
\EQs
h_\parallel^\nth{1} 
&=&
\Dinvx[\v,\ad(\e_x)h_\perp^\nth{1}]_\parallel
\nonumber\\
&=&
[\ad(\e_x)\inv\v_{xx},\v]_\parallel
-\frac{1}{2}[\ad(\e_x)\inv\v_{x},\v_x]_\parallel
+[\ad(\e_x)\inv\v,[\ad(\e_x)\v,\v_x]]_\parallel
\nonumber\\&&
+ [\v,[\ad(\e_x)\inv\v_x,\ad(\e_x)\v]]_\parallel
+\frac{1}{8}[\v,[\v, 3[\ad(\e_x)\v,\v]_\parallel 
+ [\ad(\e_x)\v,\v]_\perp ]]_\parallel
\nonumber\\&&
-\frac{1}{8}[\ad(\e_x)\inv[\ad(\e_x)\v,\v]_\perp,
[\ad(\e_x)\v,\v]_\perp]_\parallel , 
\label{hpar1st}
\endEQs
\EQs
\Ham\supth{1} 
&=&
-\langle h_\parallel^\nth{1},\e_x\rangle
= -\Dinvx\langle\v,\adsq(h_\perp^\nth{1})\rangle
\nonumber\\
&=&
-\frac{1}{2}|\v_x|^2 +\frac{1}{3}\langle\v_x,[\ad(\e_x)\v,\v]_\perp\rangle
+\frac{1}{8}|[\ad(\e_x)\v,\v]_\parallel|^2 . 
\label{ham1st}
\endEQs

The derivation of \eqref{w1st} is simple:
The local part of $\tilde\Jop$ on \eqref{hperp0th} yields
\EQ
\Dx h_\perp^\nth{0} +[\ad(\e_x)\v,h_\perp^\nth{0}]_\perp
=\v_{xx} + [\ad(\e_x)\v,\v_x]_\perp , 
\endEQ
while the nonlocal part of $\tilde\Jop$ applied to \eqref{hperp0th} 
reduces to
\EQ
[\ad(\e_x)\v,h_\parallel^\nth{0}]
=-\frac{1}{2}[\ad(\e_x)\v,[\ad(\e_x)\v,\v]_\parallel] . 
\endEQ

For the derivation of \eqref{hperp1st},
the following identity will be needed:
\EQ\label{adbrackid}
[A,(B,C)]_\parallel
- [C,(B,A)]_\parallel
= [B,(A,C)-(C,A)]_\parallel
\endEQ
which is a consequence of the Jacobi identity on the Lie bracket
$[\cdot,\cdot]$
combined with properties \eqrefs{adsymm}{adsymmperp} of the bracket
$(\cdot,\cdot) := [\ad(\e_x)\cdot,\cdot]$, 
holding for $\m$-valued functions $A,B,C$. 
To proceed, it is convenient to split up the local and nonlocal parts of
$\tilde\Hop$ on \eqref{w0th}. 
The nonlocal part consists of, via \eqref{adsymm}, 
\EQ
-\ad(\e_x)\inv
[\ad(\e_x)\v,\Dinvx[\ad(\e_x)\v,\ad(\e_x)\inv\varpi_\perp^\nth{1}]_\parallel]
= [\v,\Dinvx[\v,\varpi_\perp^\nth{1}]_\parallel]
\endEQ
which yields three terms:
First,
\EQ 
[\v,\Dinvx[\v,\v_{xx}]_\parallel]
=\ad(\v)[\v,\v_x]_\parallel
\endEQ
follows from the elementary identity $\Dx[\v,\v_x]=[\v,\v_{xx}]$. 
Second, 
\EQ
[\v,\Dinvx[\v,[\ad(\e_x)\v,\v_x]_\perp]_\parallel]
=\ad(\v)\Dinvx[\v,(\v,\v_x)]_\parallel
=\frac{1}{3}\ad(\v)[\v,(\v,\v)]_\parallel
\endEQ
is obtained from the Lie bracket relation \eqref{commpar}
followed by the identity 
$\Dx[\v,(\v,v)]_\parallel = 3[\v,(\v,\v_x)]_\parallel$
arising from \eqref{adbrackid}. 
Third, 
\EQ
-\frac{1}{2}[\v,\Dinvx[\v,[\ad(\e_x)\v,[\ad(\e_x)\v,\v]_\parallel]]_\parallel]
=-\frac{1}{4}\ad(\v)\Dinvx[(\v,\v)_\parallel, (\v,v)]_\parallel=0
\endEQ
arises from the same Lie bracket relation 
combined with \eqref{adbrackid}. 
Next the local parts of $\tilde\Hop$ on \eqref{w0th}
are straightforwardly given by expanding out 
\EQ
\adsq\inv(\Dx\varpi_\perp^\nth{1})
\endEQ
and 
\EQ
-\ad(\e_x)\inv[\ad(\e_x)\v,\ad(\e_x)\inv\varpi_\perp^\nth{1}] . 
\endEQ
Then \eqref{hperp1st} directly follows by combining these terms
with the previous ones. 

The derivation of \eqref{hpar1st} is similar. 

Now the $+1$ flow is given by 
$\v_t=h_\perp^\nth{1}+ h_\perp^\nth{0}$
which is equivalent to the evolution equation \eqref{vteq}
with $h_\perp=h_\perp^\nth{0}$ and $\varpi_\perp=\varpi_\perp^\nth{1}$.
This equation can be written out in a fairly short explicit form
by means of the brackets $[\cdot,\cdot]$ and $(\cdot,\cdot)$
together with the following linear maps
\EQ
\ad(v)_\perp := [\v,\cdot]_\perp ,\quad
\ad^\perp(v):= \ad(\e_x)\inv[\ad(\e_x)\v,\cdot]_\perp
\endEQ
defined on the vector space $\m_\perp \simeq\h_\perp$. 

{\bf Proposition~5.6: }{\it
\EQs
\v_t 
&=&
\v_x 
+ \adsq\inv(\v_{xxx})
- \adsq\inv(\frac{1}{2} (\v,(\v,\v)_\parallel) -(\v_x,\v)_\perp)_x
-(\adsq\inv(\v_{xx}),\v)_\perp
\nonumber\\&&\quad
-\ad(\v)_\perp[\v,\v_x+\frac{1}{3}(\v,\v)_\perp]_\parallel
+\frac{1}{2}\ad^\perp(\v)[\v,(\v,\v)_\parallel]
-\ad^\perp(\v)^2 \v_x
\label{mkdvflow}
\endEQs
is the $+1$ flow in the hierarchy \eqref{vevolhierarchy}. 
Up to a convective term, 
this evolution equation describes a $H^*_\parallel$-invariant
mKdV soliton equation on the $\m_\perp$-valued flow variable $\v$,
with two compatible Hamiltonian structures
\EQ
\v_t-\v_x 
= \tilde\Hop(\delta\ham{H}/\delta\v)
= \tilde\Eop(\delta\ham{E}/\delta\v) 
\endEQ
given by 
\EQ
\ham{H}
=\int( -\frac{1}{2}|\v_x|^2 +\frac{1}{3}\langle\v_x,(\v,\v)\rangle 
+\frac{1}{8}|(\v,\v)_\parallel|^2 )dx ,\quad
\ham{E}=-\int\langle\v,\adsq(\v)\rangle dx
\endEQ
where 
\EQ
\tilde\Eop:=\tilde\Rop\tilde\Hop 
= -(\ad(\e_x)\inv\Kop)^2 . 
\endEQ
}

The mKdV flow equation \eqref{mkdvflow}
corresponds to a geometrical motion of the curve $\map(t,x)$
obtained through the identifications 
\eqref{idents}, \eqref{princnor}, \eqref{perpparflows}
arising from $\m \leftrightarrow T_\map M$
as defined via a $H$-parallel moving frame along $\map(t,x)$. 
Thus, in terms of the principal normal variable \eqref{princnor},
the mKdV flow is given by 
\EQs
h_\perp 
&=&
\adsq\inv(\nu_x)
= \adsq\inv(\frder_x\nu-[\u,\nu])
= \adsq\inv\frder_x\nu -\adsq\inv[\nu,\adsq\inv[\nu,\e_x]]
\nonumber\\
&=&
\adsq\inv\frder_x\nu -(\adsq\inv\ad(\nu))^2 \e_x
\endEQs
and 
\EQs
h_\parallel 
&=&
\frac{1}{2}[\adsq\inv(\nu),\adsq\inv(\ad(\e_x)\nu)]_\parallel
=-\frac{1}{2}[\adsq\inv(\nu),[\adsq\inv(\nu),\e_x]]_\parallel
\nonumber\\
&=&
-\frac{1}{2}( \ad(\adsq\inv(\nu))^2 \e_x )_\parallel
\endEQs
where, recall, 
$\nu \leftrightarrow N= \covder{x}\mapder{x}$, 
$\frder_x\nu \leftrightarrow \covder{x} N =\covder{x}^2\mapder{x}$, 
$\e_x \leftrightarrow X=\mapder{x}$. 
Also note
\EQ
\adsq \leftrightarrow -\ad_x(X)^2 = -\ad(\mapder{x})^2:= \adsq_\map
\endEQ
which represents an invertible linear map on $T_\map M$.  
Hence 
\EQ
h_\perp +h_\parallel =\e_t
\leftrightarrow
\adsq_\map\inv\covder{x}N
-(\adsq_\map\inv\ad_x(N))^2 X 
-\frac{1}{2}(\ad_x(\adsq_\map\inv N)^2 X)_\parallel
=\mapder{t}
\endEQ
yields the following $G$-invariant geometrical equation of motion 
\EQ\label{mkdvmap}
\mapder{t} 
= 
\adsq_\map\inv\covder{x}^2\mapder{x} 
-(\adsq_\map\inv\ad_x(\covder{x}\mapder{x}))^2 \mapder{x}
-\frac{1}{2}(
\ad_x(\adsq_\map\inv\covder{x}\mapder{x})^2 \mapder{x}
)_\parallel ,\quad
|\mapder{x}|_g =1
\endEQ
which will be called a {\it non-stretching mKdV map} on $M=G/H$. 

{\it Remark: }
It is important to emphasize that 
the mKdV evolution equation \eqref{mkdvflow}
and the associated geometric map equation \eqref{mkdvmap}
are bi-Hamiltonian integrable systems that describe
multicomponent generalizations of the scalar mKdV equation. 
In particular, the evolution equation \eqref{mkdvflow}
possesses a hierarchy of higher-order commuting symmetries
and higher-order conserved densities 
corresponding to 
the vector fields \eqref{vhperphierarchy} and Hamiltonians \eqref{vham}
for $n\geq 1$. 

Both the mKdV map \eqref{mkdvmap} and evolution equation \eqref{mkdvflow}
greatly simplify in the case when the linear map
$\adsq =-\ad(\e_x)^2$ is a multiple of the identity, 
\EQ\label{adsqdiag}
\adsq =\chi\id_\perp ,\quad
\chi=\const
\endEQ
on the vector spaces $\m_\perp\simeq\h_\perp$.
From Proposition~2.1,
 note that \eqref{adsqdiag} implies
the Lie bracket relations
\EQ\label{perpbrack}
[\h_\perp,\h_\perp] \subseteq \h_\parallel ,\quad
[\m_\perp,\m_\perp] \subseteq \h_\parallel ,\quad
[\h_\perp,\m_\perp] \subseteq \m_\parallel , 
\endEQ
so consequently 
\EQ
[\cdot,\cdot]_\perp = (\cdot,\cdot)_\perp = 0 ,\quad
\ad(\v)_\perp\cdot = \ad^\perp(\v)\cdot =0 ,
\quad\text{on $\m_\perp\simeq\h_\perp$,}
\endEQ
while
\EQ
\ad^\perp(\v)\cdot =\adsq\inv\ad(\adsq(\v))_\perp\cdot = \ad(\v)\cdot\ ,
\quad\text{on $\m_\parallel$.}
\endEQ
This implies, in addition, the identity
\EQ
(A,(B,C)) - (C,(B,A)) =\chi[B,[C,A]]
\endEQ
holding for functions $A,B,C$ with values in $\m_\perp\simeq\h_\perp$.

As a result, when property \eqref{adsqdiag} holds, 
the mKdV evolution equation \eqref{mkdvflow} can be shown to reduce to 
\EQ\label{simpmkdvflow}
\v_t-\chi\v_x
= \v_{xxx} -\frac{3}{2} (\v_x,(\v,\v))
= \tilde\Hop(\delta\Ham\supth{1}/\delta\v) 
= \tilde\Eop(\delta\Ham\supth{0}/\delta\v) 
\endEQ
with a factor $\chi$ absorbed into the time derivative,
and 
$\Ham\supth{0}=-\frac{1}{2}|\v|^2$, 
$\Ham\supth{1}=-\frac{1}{2}|\v_x|^2+\frac{1}{8}|(\v,\v)|^2$,
where the bi-Hamiltonian structure becomes
\EQ
\tilde\Hop=\chi\inv\Dx-[\v,\Dinvx[v,\cdot\ ]] ,\quad
\tilde\Jop=\Dx-(\v,\Dinvx(\v,\cdot\ )) ,\quad
\tilde\Eop=\tilde\Hop\tilde\Jop\tilde\Hop ,
\endEQ
and where the recursion operator is given by 
\EQs
\tilde\Rop = \tilde\Hop\tilde\Jop
&=&
\chi\inv\Big( 
\Dx^2 -\chi[\v,[\v,\cdot\ ]] -(\v,(\v,\cdot\ ))
+ \chi[\v,\Dinvx[\v_x,\cdot\ ]] 
\nonumber\\&&\qquad
-(\v_x,\Dinvx(\v,\cdot\ ))
+ [\v,\Dinvx[\v,(\v,\Dinvx(\v,\cdot\ ))]] 
\Big)
\endEQs
(while the adjoint operator $\tilde\adRop$ is given by simply switching
the brackets $\sqrt{\chi}[\cdot,\cdot]$ and $(\cdot,\cdot)$ in $\tilde\Rop$). 

The mKdV map equation \eqref{mkdvmap} similarly simplifies to 
\EQ\label{simpmkdvmap}
\mapder{t} 
= \covder{x}^2\mapder{x} 
-\frac{3}{2}\chi\inv\ad_x(\covder{x}\mapder{x})^2 \mapder{x} . 
\endEQ

These results \eqsref{simpmkdvflow}{simpmkdvmap}
encompass the two versions of 
mKdV vector evolution equations and curve flows 
in $M=SO(N+1)/SO(N),SU(N)/SO(N)$ derived in \cite{sigmapaper},
as well as the their complex-valued generalizations obtained
from $M=SO(N+1),SU(N)$ in \cite{imapaper}. 
In particular the scalar mKdV equation and its bi-Hamiltonian structure
is given by \eqsref{simpmkdvflow}{simpmkdvmap} in the case $M=S^2$.

\subsection{Remarks on higher-order flows}

Compared to the mKdV flow \eqref{mkdvflow},
the $+n$ flow in the hierarchy \eqref{vevolhierarchy} for $n\geq 2$
is a mKdV evolution equation of higher order $2n+1$ on the flow variable $\v$.
The corresponding curve flows on $\map(t,x)$ in $M=G/H$ 
are geometrically described by higher-order versions of 
the non-stretching mKdV map equation \eqref{mkdvmap}.
All of these flows and associated geometric map equations
are bi-Hamiltonian integrable systems given by a 
group-invariant multicomponent generalization of 
the scalar mKdV soliton hierarchy coming from curve flows 
in two-dimensional constant curvature spaces, $M=S^2$ 
(cf section~1). 
Like in the scalar case, 
each flow \eqref{vevolhierarchy} turns out to be 
an explicit local polynomial expression in $\v$ and its $x$ derivatives,
for which a general proof should be possible by applying 
the locality results established in \cite{Sergyeyev}. 

There is a direct differential geometric interpretation of 
the flows \eqref{vevolhierarchy}, 
stemming from the geometrical meaning of the flow variable 
$\v=\adsq\inv(\e\hook N)$ 
given by the principal normal vector
$N=\covder{x}\mapder{x}$ along the curve $\map$
in an $H$-parallel moving frame. 
Because of the $H$-parallel property (cf Proposition~3.6),
the components of $\v$ represent differential covariants of $\map$
relative to the equivalence group $\equivH \subset H^*$ of the framing. 
More precisely, 
these components are invariantly determined by the curve $\map$
up the covariant action of the ($x$-independent) group $\equivH$ 
in the vector space $\m_\perp \subset \m\simeq_G T_x M$
containing $\v$. 
In contrast, for the case $M=S^2$,
there is essentially a unique framing 
and the equivalence group $\equivH$ comprises only discrete reflections 
about the unit tangent vector $\mapder{x}$
in the tangent plane $T_x M \simeq \Rnum^{2}$.
In this case $\v$ represents a differential invariant of $\map$, namely 
the classical curvature invariant. 
The results in Theorem~5.5 thus show that, 
in a generalization from $M=S^2$ to symmetric spaces $M=G/H$, 
differential covariants of the curve $\map$ give a proper setting 
for a flow variable that geometrically encodes 
a natural bi-Hamiltonian structure.

\subsection{SG flows}

In addition to all the local polynomial flows \eqref{vevolhierarchy}
in the hierarchy given in Theorem~5.5,
there is a nonlocal, non-polynomial flow arising from the kernel of 
the recursion operator \eqref{recursop} as follows. 
Consider the vector field $h_\perp\cdot\partial/\partial\v$ 
defined by 
\EQ
0=\varpi_\perp=\tilde\Jop(h_\perp)
\endEQ
which thus gets mapped into $\tilde\Rop(h_\perp)=0$ 
by the recursion operator. 
This will be called the $-1$ flow,
with $h_\perp$ assigned zero scaling weight under the mKdV scaling group
$x\rightarrow \lambda x$, $\v\rightarrow \lambda\inv \v$. 
Geometrically, from relation \eqref{flowprincnor}, 
$\varpi_\perp=\e\hook\covder{t}\mapder{x}=0$
immediately determines the underlying curve flow on $\map$
to be the $G$-invariant equation of motion 
\EQ\label{wavemap}
\covder{t}\mapder{x} =0 ,\qquad |\mapder{x}|_g =1
\endEQ
which is recognized as being a {\it non-stretching wave map} on $M=G/H$,
with $t$ and $x$ interpreted as light cone coordinates for the curve flow. 

The corresponding evolution equation \eqref{vteq} 
induced on the flow variable $\v$ is given by 
$h_\perp=h_\perp^\nth{-1}$ such that $\varpi=\tilde\Jop(h_\perp^\nth{-1})=0$,
so hence $\v$ satisfies the $-1$ flow equation 
\EQ\label{vsgflow}
\v_t = h_\perp^\nth{-1} ,\quad
\Dx h_\perp^\nth{-1} +[\ad(\e_x)\v,
h_\perp^\nth{-1}- \Dinvx[\ad(\e_x)\v,h_\perp^\nth{-1}]_\parallel]_\perp 
=0
\endEQ
describing a nonlocal flow. 
The structure of this equation looks simpler if we work in terms of
the flow-vector variable $h_\perp+h_\parallel=\e_t=e\hook\mapder{t}$, 
and the $x$-connection variable $\u=\ad(\e_x)\v=\conx_x=\conx\hook\mapder{x}$,
which obey the frame structure equations \eqsref{uteq}{hparwpareq}
in Lemma~4.2. 
Write 
\EQ
h^\nth{-1} := h_\perp^\nth{-1} + h_\parallel^\nth{-1}
\endEQ
and note 
\EQ\label{tconx}
\varpi^\perp+\varpi^\parallel = \conx_t =\conx\hook\mapder{t} =0
\endEQ
where 
$\varpi^\perp = -\ad(\e_x)\varpi_\perp=0$
and 
$\varpi^\parallel = -\Dinvx[\u,\varpi^\perp]_\parallel=0$
due to $\varpi_\perp=0$. 

{\bf Proposition~5.7: }{\it
The $-1$ flow equation \eqref{vsgflow} is equivalent to 
the nonlocal evolution equation
\EQ\label{sguflow}
\u_t = \ad(\e_x) h^\nth{-1}
\endEQ
with
\EQ\label{sguheq}
\Dx h^\nth{-1} =-[\u,h^\nth{-1}]
\endEQ
whose geometrical content expresses 
the vanishing of the connection \eqref{tconx}
in the flow direction $\mapder{t}$. 
}

This flow \eqsref{sguflow}{sguheq} possesses the conservation law
\EQ
0=\Dx|h^\nth{-1}|^2 ,\quad
|h^\nth{-1}|^2:= 
-\langle h_\perp^\nth{-1},h_\perp^\nth{-1}\rangle
- \langle h_\parallel^\nth{-1},h_\parallel^\nth{-1}\rangle
= |\mapder{t}|_g^2 , 
\endEQ
geometrically corresponding to 
\EQ
0=\covder{x}|\mapder{t}|_g^2
\endEQ
admitted by the wave map equation \eqref{wavemap}. 
Consequently, a conformal scaling of $t$ can be used to put
\EQ\label{sghnorm}
1= |h^\nth{-1}| = |\mapder{t}|_g
\endEQ
whence the $-1$ flow is equivalent to a flow with unit speed. 

In the examples $M=SO(N+1)/SO(N),SU(N)/SO(N)$ and $M=SO(N+1),SU(N)$, 
the conformally scaled $-1$ flow given by equations \eqsref{sguflow}{sghnorm} 
was shown \cite{sigmapaper,imapaper}
to admit an algebraic reduction yielding a hyperbolic equation of
local non-polynomial form $\v_{tx}=f(\v,\v_t)$
on the flow variable $\v=\ad(\e_x)\inv\u$.
Here a related hyperbolic equation will be derived that directly generalizes
the SG soliton equation from the scalar case $M=S^2$ to $M=G/H$,
without any reductions being introduced. 

To proceed, 
observe that equation \eqref{sguheq} has the geometric formulation
\EQ\label{sgheq}
\frder_x h^\nth{-1} =0
\endEQ
where 
\EQ\label{sgcovder}
\frder_x= \Dx +[\u,\cdot\ ]
\endEQ
is a covariant derivative operator (cf Proposition~4.4)
associated with the $x$-connection in the flow. 
Now consider a local gauge transformation on the $x$-connection,
given by a smooth function 
$\q: \map\subset M=G/H \rightarrow H^*\simeq \Ad(H)$, 
such that the operator \eqref{sgcovder} is flattened, 
$\frder_x\rightarrow \q\frder_x\q\inv=\Dx$. 
This has the geometrical meaning that the transformed $x$-connection vanishes
\EQ\label{gaugedu}
\u\rightarrow \q\u\q\inv -\q_x\q\inv :=\tilde\u =0
\endEQ
giving a linear differential equation $\q_x =\q \u$ that determines
$\q$ in terms of $\u$. 
For any such $\q$, it then follows that 
the transformed flow-vector is constant
\EQ\label{gaugedh}
h\supth{-1} \rightarrow \q h\supth{-1} \q\inv :=\tilde h ,\quad
\Dx \tilde h =0
\endEQ
where $\tilde h$ is some constant $\m$-valued vector
with unit norm due to property \eqref{sghnorm}.

As a consequence, equation \eqref{sgheq} is solved by 
\EQ\label{sgpot}
\u =\q\inv \q_x ,\quad
h\supth{-1} = \q\inv \tilde h\q
\endEQ
in terms of a $H^*$-valued potential $\q$. 
Substitution of these expressions into the $-1$ flow equation \eqref{sguflow}
now leads to a main result. 

{\bf Theorem~5.8: }{\it
The $-1$ flow on $\v=\ad(\e_x)\inv\u=\ad(\e_x)\inv(\q\inv\q_x)$ 
is given by a hyperbolic equation 
\EQ\label{sgflow}
\q_{tx} -\q_t \q\inv \q_x 
= \q[\e_x,\q\inv \tilde h\q]
= [\q \e_x\q\inv,\tilde h]\q
\endEQ
for the $H^*$-valued potential $\q$,
with $\Dx \tilde h=0$ and $|\tilde h|_\m=1$. 
}
 
This flow depends on the pair of constant unit vectors 
$\e_x,\tilde h$ in $\m$.
Without loss of generality, 
under a rigid ($x$-independent) transformation 
$\tilde h \rightarrow \tilde\q \tilde h \tilde\q\inv$
if necessary
(which preserves \eqref{sgpot}),
the vector $\tilde h$ can be taken to belong to 
the same Weyl chamber $a_*(\m)$ that contains the vector $\e_x$
in a fixed maximal abelian subspace $\a\subset\m$,
whereby the flow \eqref{sgflow} is parametrized by 
unit-norm elements in the Weyl chamber $a_*(\m)$ in $\m$
(cf Proposition~3.2). 
Note the invariance group of the flow will be $\equivH$ iff 
$\tilde h=\e_x$. 

A natural Hamiltonian structure can be derived for 
the hyperbolic flow equation \eqref{sgflow}
starting from the scaling formula \eqref{vham} extended to $n=-1$. 
Let
\EQ
\Ham\supth{-1} :=\langle\e_x,h_\parallel^\nth{-1}\rangle 
=\langle\e_x,h\supth{-1}\rangle 
=\langle\e_x,\q\inv \tilde h \q\rangle . 
\endEQ
Its variational derivative with respect to $\q$ is defined by 
$\delta\Ham\supth{-1} 
=\tr( \delta\q (\delta\Ham\supth{-1}/\delta\q) )
= -\langle\q\inv\delta\q,(\delta\Ham\supth{-1}/\delta\q)\q\rangle$,
which yields
\EQ\label{sgvarham}
-(\delta\Ham\supth{-1}/\delta\q)\q
= [\e_x,\q\inv \tilde h \q]
= \ad(\e_x)h\supth{-1} . 
\endEQ

{\bf Theorem~5.9: }{\it
The hyperbolic flow equation \eqref{sgflow}
has the Hamiltonian structure
\EQ\label{sghamflow}
\q\inv( \q_{tx} -\q_t \q\inv \q_x )\q\inv
= -\delta\Ham\supth{-1}/\delta\q
\endEQ
or equivalently
\EQ\label{sghamflowalt}
\q_t \q\inv =-\Dinvx(\q (\delta\Ham\supth{-1}/\delta\q))
\endEQ
as an evolutionary flow,
where $\Dinvx$ is a Hamiltonian operator 
on the $x$-jet space of $\h$-valued functions. 
}

This Hamiltonian structure has an equivalent formulation in terms of
the flow variable $\v=\ad(\e_x)\inv\u$ by means of the variational relations
\EQs
&&
\delta\Ham\supth{-1} 
=-\langle\delta\u,\delta\Ham\supth{-1}/\delta\u\rangle
=-\langle\delta\v,\delta\Ham\supth{-1}/\delta\v\rangle , 
\\
&&
\delta\u=\ad(\e_x)\delta\v=\delta(\q\inv\q_x) =\frder_x (\q\inv\delta\q) , 
\label{varu}
\endEQs
combined with the variational derivative \eqref{sgvarham},
where 
$\delta\Ham\supth{-1}/\delta\u$ 
and 
$\delta\Ham\supth{-1}/\delta\v$ 
are $\h_\perp$- and $\m_\perp$-valued respectively. 
Note, since $(\delta\u)_\parallel=0$, 
the relation \eqref{varu} implies
$(\q\inv\delta\q)_\parallel=-\Dinvx[\u, (\q\inv\delta\q)_\perp]_\parallel$
and hence
\EQ
\delta\u 
= \Dx (\q\inv\delta\q)_\perp +[\u,
(\q\inv\delta\q)_\perp -\Dinvx[\u, (\q\inv\delta\q)_\perp]_\parallel ]_\perp
= \Kop(\q\inv\delta\q)_\perp 
\endEQ
where $\Kop$ is the operator \eqref{Kop}. 

{\bf Proposition~5.10: }{\it
The $-1$ flow equation \eqsref{sguflow}{sguheq} has the Hamiltonian structure
\EQ\label{sguhamflow}
\u_t =\Hop ( \delta\Ham\supth{-1}/\delta\u )
\endEQ
where $\Ham\supth{-1}=\langle\e_x,\q\inv \tilde h\q\rangle$
is a nonlocal function of $\u$ determined by $\q_x =\q \u$. 
Equivalently, 
the $-1$ flow equation \eqref{vsgflow} on $\v=\ad(\e_x)\inv\u$
is given by 
\EQ\label{sgvhamflow}
\v_t =\tilde\Hop( \delta\Ham\supth{-1}/\delta\v )
\endEQ
with $\q$ expressed in terms of $\v$ through $\q_x =\q[\e_x,\v]$. }

As a consequence, the $-1$ flow on $\v$ thus defines 
a nonlocal Hamiltonian vector field
\EQ
h_\perp^\nth{-1} :=(\q\inv \tilde h \q)_\perp
=\tilde \Hop(\delta\Ham\supth{-1}/\delta\v)
\endEQ
and likewise
\EQ
h^\perp_\nth{-1} := \ad(\e_x) h_\perp^\nth{-1} 
= [\e_x, \q\inv \tilde h \q] = \Hop(\delta\Ham\supth{-1}/\delta\u)
\endEQ
is a corresponding nonlocal Hamiltonian vector field
given by the $-1$ flow on $\u$. 

{\it Remarks: }
The group-valued hyperbolic equation 
\eqref{sghamflow} 
provides a multicomponent generalization of the scalar SG equation
along with its Hamiltonian structure,
arising from curve flows \eqref{wavemap} in $M=G/H$. 
In particular, for $M=S^2$
if $H^* =SO(2)\simeq U(1)$ is viewed as the group of abelian unitary matrices
then $\q=\exp(\i\theta)$ can be identified as a $U(1)$-valued potential
satisfying the SG equation 
$\theta_{tx} =\sin(\theta-\tilde\theta)$
with the Hamiltonian 
$\Ham\supth{-1}=\i\Im(\exp(\i\theta)\exp(-\i\tilde\theta))$,
$\tilde\theta=$constant. 
From Theorems~5.8 and~5.9 and Proposition~5.10, 
the generalization of this soliton equation 
in the multicomponent (nonabelian) case
describes an integrable SG system for which the associated $-1$ flow
in nonlocal evolutionary form \eqref{sghamflowalt}
possesses a hierarchy of
higher-order commuting symmetries given by 
the vector fields \eqref{vhperphierarchy} 
and higher-order conserved densities given by 
the Hamiltonians \eqref{vham} 
for $n\geq 0$. 
Moreover, these hierarchies can be expressed directly in terms of
the potential $\q$ via the relation 
$\v=\ad(\e_x)\inv\u = \ad(\e_x)\inv(\q\inv\q_x)$.

\section{Concluding Remarks}

The main results in this paper
on bi-Hamiltonian hierarchies of geometric curve flows
can be directly extended to flat Riemannian symmetric spaces 
$\m=\g/\h\simeq \Rnum^n$
viewed as infinitesimal Klein geometries. 
One significant geometrical difference will be that 
both the curvature and torsion of the Riemannian connection $\covder{}$
will vanish in the Cartan structure equations for the framing of
the curve flows $\map(t,x)$ in $\Rnum^n$. 
This change does not affect the encoding of bi-Hamiltonian operators,
which will be the same Hamiltonian cosymplectic and symplectic operators
as derived in curved symmetric spaces $M=G/H$, 
but it does lead to a simpler recursion structure 
for the resulting integrable hierarchy of group-invariant flow equations
induced on the moving frame components of the principal normal vector
along the curves. 
In particular, 
the higher-order mKdV flows in each hierarchy for $M=G/H$ 
(when $\covder{}$ has non-zero curvature)
are linear combinations of the ones arising in $\m=\g/\h$
(when $\covder{}$ is flat). 

A further extension of the main results in the present paper
arises for Riemannian symmetric spaces $M=G/H \simeq K$ 
given by any compact semisimple Lie group $K$, 
where $G=K\times K$ is a product group
and $H\simeq K$ is a diagonal subgroup. 
Since $M$ here is a group manifold, 
there turns out to be an algebraic Hamiltonian operator that is compatible
with the cosymplectic and symplectic operators encoded in 
the Cartan structure equations for the framing of curve flows $\map(t,x)$. 
This leads to an enlarged bi-Hamiltonian hierarchy of geometric curve flows,
containing group-invariant NLS flows in addition to mKdV flows. 
The same result extends to the flat spaces $\m=\g/\h \simeq \alg{k}$
for any compact semisimple Lie algebra $\alg{k}$
(viewed as an infinitesimal Klein geometry). 
Details of this will presented in a forthcoming paper. 

Enlarged bi-Hamiltonian hierarchies of geometric curve flows
will likewise arise for other symmetric spaces that carry 
extra algebraic structure, such as hermitian spaces or quaternion spaces. 
All such spaces are well-known from Cartan's classification
\cite{Helgason}.

Earlier work \cite{sigmapaper,imapaper} 
studying the four symmetric spaces 
$SO(N+1)/SO(N)$, $SU(N)/SO(N)$, $SO(N+1)$, $SU(N)$ 
geometrically accounted for all examples of
$O(N-1)$-invariant and $U(N-1)$-invariant multicomponent mKdV and SG equations
that have been found in symmetry-integrability classifications of
vector evolution equations \cite{SokolovWolf} 
and vector hyperbolic equations \cite{AncoWolf}. 
Building on these examples, 
a subsequent paper will be devoted to a full classification of
further examples of group-invariant soliton equations derived 
from geometric curve flows in all classical (non-exceptional) 
symmetric spaces.

Finally, in light of work done in \cite{ChouQu1,ChouQu2} 
deriving numerous scalar soliton equations from geometric curve flows 
in planar Klein geometries $G/H \simeq \Rnum^2$, 
a natural interesting question is whether 
the present analysis using moving parallel frames for curve flows
in Riemannian symmetric spaces 
can be broadly extended to general homogeneous spaces $M=G/H$.
Such a generalization would encompass non-Riemannian Klein geometries
and, accordingly, should yield group-invariant soliton equations
(along with their bi-Hamiltonian structure) of other than mKdV and SG type.

\subsection*{Acknowledgments} 
The author is supported by an N.S.E.R.C. grant. 
Jing Ping Wang and Jan Sanders are thanked
for valuable conversations at the time of the BIRS meeting on
``Differential Invariants and Invariant Differential Equations''
which stimulated the research in this paper.

\end{document}